\documentclass[amsmath,amssymb,superscriptaddress,nofootinbib]{revtex4}

\usepackage{graphicx}
\usepackage{dcolumn}
\usepackage{bm}
\usepackage[small,bf]{caption}

\newcommand{\svec}[1]{\bm{#1}}
\newcommand{\ivec}[1]{\vec{#1}}
\newcommand{\kdag}{\svec{k}^\dagger} 
\newcommand{\scriptst}{{ }}
\newcommand{\sigmamat}{{\svec\sigma_{\sigma' \sigma}}}

\begin{document}


\title{Self-Consistent Description of  Multipole Strength in Exotic Nuclei I:
Method}


%
%

 \author{J. Terasaki}
 \email{jterasak@physics.unc.edu}
 \affiliation{Department of Physics and Astronomy, CB3255, University of North
  Carolina,  Chapel Hill, NC 27599-3255}
 \affiliation{Department of Physics and Astronomy, University of Tennessee,
  Knoxville, TN 37996}
 \affiliation{Physics Division, Oak Ridge National Laboratory, P.O.~Box 2008,
  Oak Ridge, TN 37831}
 \affiliation{Joint Institute for Heavy-Ion Research, Oak Ridge, TN 37831}

 \author{J. Engel}
 \email{engelj@physics.unc.edu}
 \affiliation{Department of Physics and Astronomy, CB3255, University of North
  Carolina,  Chapel Hill, NC 27599-3255}

 \author{M. Bender}
 \email{mbender@phy.anl.gov}
 \affiliation{Physics Division, Argonne National Laboratory, Argonne, IL 60439}

 \author{J.~Dobaczewski}
 \email{dobaczew@fuw.edu.pl}
 \affiliation{Department of Physics and Astronomy, University of Tennessee,
  Knoxville, TN 37996}
 \affiliation{Physics Division, Oak Ridge National Laboratory, P.O.~Box 2008,
  Oak Ridge, TN 37831}
 \affiliation{Joint Institute for Heavy-Ion Research, Oak Ridge, TN 37831}
 \affiliation{Institute of Theoretical Physics, Warsaw University,
  ul.~Ho\.{z}a 69, 00-681 Warsaw, Poland}

 \author{W. Nazarewicz}
 \email{witek@utk.edu}
 \affiliation{Department of Physics and Astronomy, University of Tennessee,
  Knoxville, TN 37996}
 \affiliation{Physics Division, Oak Ridge National Laboratory, P.O.~Box 2008,
  Oak Ridge, TN 37831}
 \affiliation{Institute of Theoretical Physics, Warsaw University,
  ul.~Ho\.{z}a 69, 00-681 Warsaw, Poland}

 \author{M. Stoitsov}
 \email{stoitsovmv@ornl.gov}
 \affiliation{Department of Physics and Astronomy, University of Tennessee,
  Knoxville, TN 37996}
 \affiliation{Physics Division, Oak Ridge National Laboratory, P.O.~Box 2008,
  Oak Ridge, TN 37831}
 \affiliation{Joint Institute for Heavy-Ion Research, Oak Ridge, TN 37831}
 \affiliation{Institute of Nuclear Research and Nuclear Energy, Bulgarian
  Academy of Science, Sofia 1784, Bulgaria}


\date{\today}

\begin{abstract}
We use the canonical Hartree-Fock-Bogoliubov
basis to implement a completely self-consistent
quasiparticle-random-phase approximation with arbitrary Skyrme energy density
functionals and density-dependent pairing functionals.  The point
of the approach is to accurately describe multipole
strength functions in spherical
even-even nuclei, including weakly-bound drip-line systems.  We
describe the method and carefully test its accuracy, particularly in handling
spurious modes.  To illustrate our approach, we calculate
isoscalar and isovector
monopole, dipole, and quadrupole
 strength functions
in several  Sn isotopes, both in the stable region and at the drip lines.
\end{abstract}

\maketitle

\section{\label{sec:intro}Introduction}

The study of nuclei far from stability is an increasingly important part of
nuclear physics \cite{NSAC,NUPECC}.  As radioactive beams allow more
experiments on these nuclei, theoretical modeling is changing in 
significant
ways.  New ideas and progress in computer technology have allowed nuclear
theorists to understand bits and pieces of nuclear structure quantitatively
\cite{theory}.  Short-lived exotic nuclei offer unique tests of those aspects
of our developing many-body theories that depend on neutron
excess~\cite{dob98}. The major challenge is to predict or describe in detail
exotic new properties of nuclei far from the stability valley, and to
understand the origins of these properties.

For medium-mass and heavy nuclei, an important goal is obtaining a universal
energy-density functional, which will be able to describe static and dynamic
properties of finite nuclei and extended nucleonic matter with arbitrary
neutron-to-proton ratio.  Self-consistent methods based on density-functional
theory are already sophisticated enough to allow precise analysis of
ground-state properties (e.g.\ binding energies) in heavy nuclei
\cite{gor02,gor03,sto03}.  They can also help describe nuclear decays and
excited states. Their predictions for collective excitations as we approach the
neutron drip line are especially interesting.  What happens to low- and
high-frequency multipole modes when the neutron excess is unusually large?

To address these questions we use the quasiparticle random-phase approximation
(QRPA), a powerful tool for understanding both low-lying vibrational states and
giant resonances \cite{rin80}.  The QRPA is a microscopic approach that is
nevertheless simple enough to allow ``no-core'' calculations.  The
approximation, which should  be good for collective vibrations as long as their
amplitudes are small, is especially effective in conjunction with Skyrme energy
functionals.  Our work is a part of a broad program to test and improve these
functionals, which thus far have been fitted mainly to ground-state
observables, by applying them to collective excitations, particularly near the
drip line.  This paper lays out our approach and evaluates its accuracy.  For
these purposes we restrict ourselves to a single Skyrme functional, SkM$^*$.  A
forthcoming study  will examine the performance of Skyrme functionals more
generally.

The QRPA is a standard method for describing collective excitations in
open-shell superconducting nuclei with stable mean-field solutions, either
spherical or deformed. What is not standard, and at the same time is extremely
important for weakly bound nuclei, is the treatment of the particle continuum.
Continuum extensions of the random phase approximation (RPA) or QRPA are
usually carried out in coordinate space, facilitating treatment of decay
channels and guaranteeing correct asymptotics.  Surprisingly, as we discuss
below, the rich literature on the RPA and QRPA, which includes many
coordinate-space calculations, contains few treatments of the continuum that
exploit the entire Skyrme functional in a fully self-consistent way.

To avoid confusion, we state what we mean by a fully self-consistent RPA or
QRPA calculation.  First, the underlying mean-field calculation must be
self-consistent in the usual sense.  Next, the residual interaction used in the
RPA or QRPA must be derived from the same force or energy functional that
determines the mean field.  An important consequence of this condition, and of
other more detailed technical conditions discussed below, is that spurious
excitations arising from symmetry breaking by the mean field have zero or
nearly zero energy, leaving the physical intrinsic excitations completely
uncontaminated by spurious motion. Finally, energy-weighted sum rules must be
satisfied to high accuracy.  We elaborate on these requirements below;
Refs.~\cite{shl02,agr03,agr04} discuss ways in which RPA calculations commonly
violate them.

The literature applying RPA or QRPA to nuclear structure is huge, and a
complete review is beyond the scope of our paper.  We do, however, present an
overview of the studies that are related in one way or another to nuclear
density functionals, self consistency, pairing, and the key issue of the
particle continuum.

The standard version of QRPA, the so-called matrix formulation, is carried out
in the configuration space \cite{row70,war87} of single-quasiparticle states. A
number of papers treat collective states in spherical nuclei in the Skyrme-RPA
and QRPA matrix formulation (see Refs. \cite{war87,col03} and references cited
therein), in which the positive-energy continuum is discretized, e.g.\ by
solving the Hartree-Fock-Bogoliubov (HFB) and QRPA equations in a
harmonic-oscillator single-particle basis.  Within this group, the first fully
self-consistent calculations that properly account for continuum effects are
those of Refs.~\cite{eng99,ben02}, in which the localized canonical basis of
coordinate-space HFB is used to calculate beta-decay rates of neutron-rich
r-process nuclei and Gamow-Teller strength distributions.  Recently, fully
self-consistent HFB+QRPA calculations have also been carried out with the
finite-range Gogny force \cite{gia03}. Unlike many previous Gogny+HFB studies
that employed a harmonic oscillator basis, Ref.~\cite{gia03} solves the HFB
equations in the eigenbasis of a Woods-Saxon potential, the particle continuum
of which is discretized by enclosing the system in a box.

Coordinate-space Green's functions as a method of implementing the RPA
through linear response were first used in Ref.~\cite{shl75} and
subsequently applied to the description of low- and high-energy nuclear
modes (see, e.g.,
Refs.~\cite{liu76,bar80,ham97,ham98,ham99,ham00,kol00,sag02,ham02,shl02,shl03}).
Many of those calculations are not realistic enough, however, because they
ignore the spin-orbit and Coulomb residual interactions in the RPA
\cite{agr03,agr04}.  Coordinate-space Green's-function QRPA was studied in
Ref.~\cite{hag01}, in the BCS approximation, with a phenomenological
Woods-Saxon average potential.  Coordinate-space HFB+QRPA for spherical
nuclei was formulated in Refs.~\cite{mat01,mat02,kha02,kha04} and applied
to excitations of neutron-rich nuclei. As in \cite{hag01}, the
Hartree-Fock (HF) field in Refs.~\cite{mat01,mat02} was approximated by a
Woods-Saxon potential.  While the calculations of Refs.~\cite{kha02,kha04}
are based on Skyrme-HFB fields, they violate full self consistency by
replacing the residual velocity-dependent terms of the Skyrme force by the
Landau-Migdal force in the QRPA, and neglecting spin-spin, spin-orbit, and
Coulomb residual interactions entirely. Within this approach, extensive
Skyrme-HF+BCS QRPA calculations of E1-strength in neutron-rich nuclei were
carried out in Refs.~\cite{gor02a,gor04}.

An alternative coordinate-representation approach, also based on Green's
functions, was formulated in Refs.~\cite{pla88,kam04} within Migdal's
finite-Fermi-systems theory.  Most of practical applications of this method,
however, involve approximations that break self consistency in one way or
another, including  the use of highly truncated pairing spaces, different
interactions in HFB and QRPA, and the so-called diagonal pairing approximation
\cite{pla88,fay94,hor96,bor95,bor96,kam98,kam01,bor03,kam04a}.  Properties of
excited states and strength functions have also been investigated within the
relativistic RPA \cite{rin01,vre01,ma02,vre02,nik02,ma03,vre04} or QRPA
\cite{paa03,paa04}.  The QRPA work employs the matrix formulation and is fully
self-consistent, since it uses the same Lagrangian in the relativistic
Hartree-Bogoliubov calculation of the ground state and in the QRPA matrix
equations, which are solved in the canonical basis.

At the present, no fully self-consistent continuum  HFB+QRPA calculations exist
in deformed nuclei.  Refs.\ \cite{nak02,nak04} studied giant resonances in
deformed nuclei within time-dependent HF theory, formulated in coordinate space
with a complex absorbing boundary condition imposed.  Symmetry-unrestricted RPA
calculations, with no pairing, were carried out in Ref.~\cite{ina04} in
a ``mixed representation'' \cite{ima03} on a Cartesian mesh in a box, while
Ref.~\cite{hag04} contains examples of BCS+QRPA calculations in the
single-particle basis of a deformed Woods-Saxon potential.

The work described in this paper is fully self-consistent: among other things
we use precisely the same interaction in the HFB and QRPA calculations so as to
preserve the small-amplitude limit of time-dependent HFB.  We formulate the
QRPA  in the canonical eigenbasis of the one-body particle-density matrix
\cite{dob96} which is calculated in the coordinate representation in a large
spherical box.  As mentioned above, the canonical basis has been used
previously to study $\beta$ decay and Gamow-Teller strength \cite{eng99,ben02};
its use in charge-conserving modes near the drip line is more challenging,
however, because of the existence of spurious states in the monopole and dipole
channels.\footnote{As far as we know, the only application of the canonical
basis to charge-conserving modes near the drip line is in the relativistic
QRPA, see e.g.\ \cite{rin03,paa03}.}  These zero-energy modes can mix with
physical states unless the QRPA equations are solved with high accuracy.  A
less precise implementation of our approach was used to calculate
neutrino-nucleus cross sections in $^{208}$Pb in Ref.\ \cite{eng03}.

This paper is organized as follows.  Section~\ref{sec:method} below presents our
approach.  In Sec.~\ref{sec:accuracy} we check the QRPA solutions carefully,
focusing on spurious modes.  Section \ref{sec:conclusion} contains the main
conclusions of our work.  Mathematical details are in two appendices, the first
of which is on the QRPA equations and the second on calculating the derivatives
of the Skyrme functionals that enter the formalism.

\section{\label{sec:method} Method}

Our first step in the self-consistent treatment of excitations is to solve the
spherical HFB equations in coordinate space (without mixing neutron
and proton quasiparticle wave functions \cite{per04}), with the
method developed in Ref.\ \cite{dob84} (see also
Refs.~\cite{dob96,dob01a,ben04}).  We can use arbitrary Skyrme functionals
in the particle-hole and pairing (particle-particle) channels.

We modify the code used in Refs.\ \cite{dob84,dob96,dob01a} so that it solves
the HFB equations with higher accuracy, which we need because the QRPA uses all
the single-quasiparticle states produced by the HFB equations, even those that
are essentially unoccupied.  Our modifications are:  (i) the use of quadruple
precision (though in solving the QRPA equations we use  double precision); (ii)
a smaller discretization length (0.05 fm); and (iii) a high
quasiparticle-energy cutoff (200 MeV) and a maximum angular momentum $j_{\rm
max}$=15/2 ($N\leq 82$) or 21/2 ($N>82$). In a 20 fm box, this cutoff
corresponds to 200--300 quasiparticle states for each kind of nucleon. We
include all these quasiparticle states in the HFB calculation because a very
large energy cutoff is essential for the accuracy of self-consistent QRPA
calculations \cite{agr03}.  Hence, the effective pairing window in our HFB
calculations is also very large, with the pairing functional fitted to
experimental pairing gaps extracted as in Ref.\ \cite{ter02} from the measured
odd-even mass differences in several Sn, Ni, and Ca isotopes. 

Next, we construct the canonical basis, the eigenstates of single-particle
density matrix $\rho$.  To avoid poor accuracy (see Ref.\ \cite{dob96}) in the
wave functions of the nearly empty canonical particle states, we do not
diagonalize $\rho$ directly in coordinate space.  Instead we construct an
intermediate basis by orthonormalizing a set of functions $\{
\varphi^{\mu}_1({\bm r}) +\varphi^{\mu}_2( {\bm r})\}$, where
$\varphi^{\mu}_1({\bm r})$ and $\varphi^{\mu}_2( {\bm r})$ are the upper and
lower components of the quasiparticle wave function with energy $E_{\mu}$
\cite{dob84}.  We use the density matrix in coordinate space to calculate the
matrix in this basis, which we then diagonalize to obtain the canonical
states.  The reason for using the sum of $\varphi^{\mu}_1({\bm r})$ and
$\varphi^{\mu}_2({\bm r})$ is that solutions of the HFB equations expressed in
the canonical basis (Eqs. (4.14) of Ref.~\cite{dob96}) are, in the new basis,
{\it guaranteed} to be numerically consistent with those of the original HFB
problem. This is because the configuration space is the same in both cases,
independent of the pairing cutoff (see  Ref.~\cite{dob04a} for a discussion
relevant to this point). Without pairing, when either $\varphi^{\mu}_1({\bm
r})$ or $\varphi^{\mu}_2( {\bm r})$ is equal to zero, our method is equivalent
to taking a certain number of HF states, including many unoccupied states.

In the canonical basis, the HFB+QRPA equations have a form almost identical to
that of the BCS+QRPA approximation, the only difference being the presence of
off-diagonal terms in the single-quasiparticle energies.  The QRPA+HFB
formalism employs more pairing matrix elements than the QRPA+BCS, however.

As noted already, full self consistency requires the use of the same
interaction in the QRPA as in the HFB approximation.  More specifically, this
means that the matrix elements that enter the QRPA equation are related to
second derivatives of a mean-field energy functional.  We describe the
densities and the form of the functional carefully in the appendices.  But we
must meet other conditions as well for QRPA calculation to be self-consistent.
Essentially all the single-particle or quasiparticle states produced by the HFB
calculation must be used in the space of two-quasiparticle QRPA excitations.
This requirement is rather stringent, so we truncate the two-quasiparticle
space at several levels and check for convergence of the QRPA solution.  First
we omit canonical-basis wave functions that have occupation probabilities
$v_i^2$ less than some small $v_{\rm crit}^2$, (or HF energies greater than
some $\varepsilon_{\rm crit}$ if there is no pairing).  Then we exclude from
the QRPA pairs of canonical states for which the occupation probabilities are
both larger than $1-v_{\rm crit}^2$.  This second cut is based on the
assumption that two-particle transfer modes are not strongly coupled to
particle-hole excitations.  In addition, if the factors containing $u_i$ and
$v_i$ in the QRPA equation --- see Eqs.\ (\ref{A_spherical}) and
(\ref{B_spherical}) --- are very small, in practice smaller than 0.01, then we
set the corresponding matrix elements equal to zero. This does not affect the
size of the QRPA space, but significantly speeds up the calculations.  For good
performance we diagonalize QRPA-Hamiltonian matrices of order $20,000 \times
20,000$ in neutron-rich Sn isotopes.

Having solved the QRPA equations, we can then calculate the strength function
\begin{equation}
S_J(E) =
\frac{1}{\pi}\sum_{k}\sum_{M=-J}^J\frac{\gamma(E_k)|\langle k|
\hat{F}_{JM}|
0\rangle|^2}{(E_k-E)^2 + \gamma^2(E_k)},
\label{strength_function}
\end{equation}
for the multipole operator $\hat{F}_{JM}$. The smoothing
width $\gamma$ is supposed to be large enough  to remove spurious
oscillations in $S_J(E)$ associated with a finite box radius $R_{\rm box}$
\cite{shl97,nak02}. A reasonable form, based
on a single-particle estimate, for the smoothing width (App. B of Ref.~\cite{shl97}),  is
\begin{eqnarray}\label{smwidth}
\gamma(E)=\left\{\begin{array}{ll}
\frac{\pi}{R_{\rm box}}\sqrt{\frac{\hbar^2(E+\lambda_{\rm n})}{2m}},
& E\geq-\lambda_{\rm n} \\0.1~{\rm MeV}, & E<-\lambda_{\rm n}
\end{array}\right.,
\end{eqnarray}
where  $\lambda_{\rm n}$ is the neutron Fermi level
and $m$ is the nucleon mass.
In deriving Eq.~(\ref{smwidth})
we assumed that the  single-proton continuum is effectively shifted up
several MeV by the Coulomb barrier. In other words, we associate the threshold energy with the neutron
Fermi level.

In all the tests below, we use the  Skyrme functional SkM$^\ast$ \cite{bar82}
and a volume pairing functional \cite{dob02} ($C^{\tilde{\rho}}(\rho_{00})$ a
constant in Eq.\ (\ref{pairing_energy_functional})).  The pairing parameter in
Eq.\ (\ref{volpair}) is $V_0=-77.5$ MeV fm$^3$.  Usually we work in a box of
radius 20 fm, though we vary this radius below to see its effects.  In several
tests we examine the weakly bound nucleus $^{174}$Sn, which is very close to
the two-neutron drip line.  In this system, the protons are unpaired and the
neutrons paired (with $\Delta_{\rm n}$=1.016\,MeV) in the HFB ground state.

\section{\label{sec:accuracy} Accuracy of solutions}
\begin{figure}[t]
\includegraphics[width=12cm]{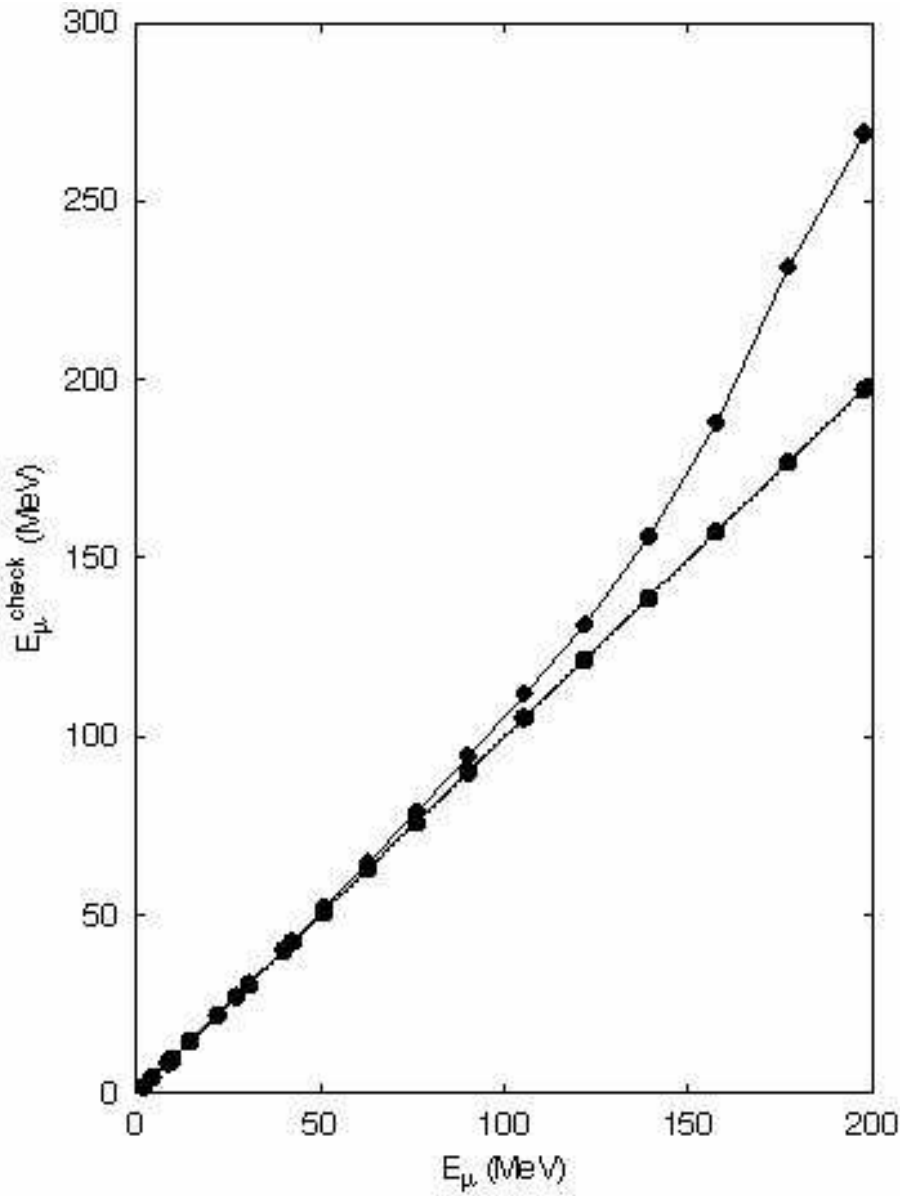}
\caption{\label{fig:ehfb_eigh11}
Neutron quasiparticle energies $E_\mu^{\rm check}$ for
$s_{1/2}$ states in $^{174}$Sn,
calculated by diagonalizing
the HFB Hamiltonian in the canonical basis, versus the quasiparticle
energies $E_\mu$ obtained by directly solving the HFB equations
in coordinate space. Standard (solid line, diamonds) and
improved (dotted line, dots) methods are used to obtain the canonical states.
See text for details.
}
\end{figure}
Benchmark tests of the HFB part of our
calculations are reported in Ref.\ \cite{dob04}. Since the
accuracy of the canonical wave functions, in which the QRPA calculations
are carried out,  strongly affects the quality of results (in particular
QRPA self consistency), we take special care to
compute them precisely.
As  discussed in Sec.~\ref{sec:method}, we obtain canonical states
by diagonalizing the
single-particle density matrix $\rho$ represented in the
orthonormalized set of functions
$\{ \varphi^{\mu}_1({\bm r})
+\varphi^{\mu}_2( {\bm r})\}$. The accuracy
of this method  is illustrated in Fig.~\ref{fig:ehfb_eigh11}, which
plots the quasiparticle energies $E_\mu^{\rm check}$, obtained
by diagonalizing the HFB Hamiltonian in the canonical basis
(Eq. (4.20) of Ref.~\cite{dob96}), versus the quasiparticle
energies $E_\mu$ obtained by solving the HFB differential
equations directly in coordinate space
(Eq. (4.10) of Ref.~\cite{dob96}).
Two sets of canonical states are used: (i) those obtained through
the procedure outlined above (dotted line) and (ii)
those obtained in the standard way
by diagonalizing
the density matrix $\rho(\bm{r},\bm{r}^\prime)$ in discretized coordinate
space (Eq. (3.24a) of Ref.~\cite{dob96}; solid line).
If the canonical basis is precisely determined, $E_\mu^{\rm check}$=$E_\mu$ and
the  two sets of $E_\mu^{\rm check}$ coincide.  Within the standard approach, however,
the high canonical energies
deviate visibly from their HFB counterparts, i.e., the accuracy of the underlying canonical wave functions 
is poor. On the other hand, the quasiparticle energies and canonical
wave functions
calculated within the modified approach introduced above
are as accurate as the original solutions to the HFB equations,
even for high-lying nearly-empty
states.
(See also Sec.~VI.D of Ref.~\cite{taj04} for a discussion relevant to this
point.)

Having examined the canonical basis, we turn to the
accuracy of the QRPA part of the calculation.
To test it, we first consider solutions related
to symmetries.
If a Hamiltonian is invariant
under a symmetry operator $\hat{P}$ and the HFB state $|\Psi\rangle$
spontaneously breaks the symmetry, then $e^{i\alpha\hat{P}}|\Psi\rangle$, with
$\alpha$ an arbitrary c-number, is
degenerate with the state $|\Psi\rangle$.  The 
QRPA equations have a spurious solution at zero energy associated with the
symmetry breaking \cite{rin80,bla86}, while all other solutions are free of the
spurious motion.
This property is important for strength functions, and gives us a way of testing
the calculations.  Since our QRPA equations, which assume spherical symmetry,
are based on mean fields that include pairing and are localized in space,
there appear
spurious states associated with particle-number nonconservation (proton and/or
neutron; 0$^+$ channel) and center-of-mass motion (1$^-$ channel).
These two cases are discussed below in
Sec.~\ref{zeroplus} and Sec.~\ref{oneminus}.
\begin{table}[b]
\caption{\label{tab:strength_number} The lowest-energy excited
$0^+$ states in $^{174}$Sn.
The second column shows the excitation energies
and the third column the squared matrix elements of the particle-number operator
between the $k$th excited state and the ground state ($k$=0).}
\parbox{7cm}{
\begin{ruledtabular}
\begin{tabular}{lll}
{\it k} &$E_k$ (MeV) & $|\langle k|\hat{N}|0\rangle|^2$ \\
\hline
1 & 0.171 &  0.120 \\
2 & 2.833 &  $0.533\times 10^{-5}$ \\
3 & 3.090 &  $0.877\times 10^{-7}$ \\
4 & 3.810 &  $0.252\times 10^{-5}$ \\
5 & 3.878 &  $0.480\times 10^{-5}$
\end{tabular}
\end{ruledtabular}
}
\end{table}

\subsection{The 0$^+$ isoscalar mode}\label{zeroplus}

In addition to the spurious state associated with nonconservation of particle-number by the HFB, the $0^+$ channel contains the important ``breathing mode''.  
In Table~\ref{tab:strength_number} we display results from a run with $v^2_{\rm
crit}$=$10^{-12}$ for neutrons and $\varepsilon_{\rm crit}=150$ MeV for protons,
resulting in the inclusion of 310 proton quasiparticle states and the same number of
neutron states, with angular momentum up to $j=21/2$.
The Table shows the QRPA energies and transition matrix elements of the
particle-number operator.  The spurious state is below 200 keV, well separated from the other states, all of which have negligible ``number-strength''.  The nonzero number strength in the
spurious state, like the nonzero energy of that state, is a measure of
numerical error.
If the space of two-quasiparticle states is smaller, with $\varepsilon_{\rm
crit}=100$ MeV and $v^2_{\rm crit}=10^{-8}$, the energy of the spurious state
and the number strength barely change.
\begin{figure}[t]
\includegraphics[width=12cm]{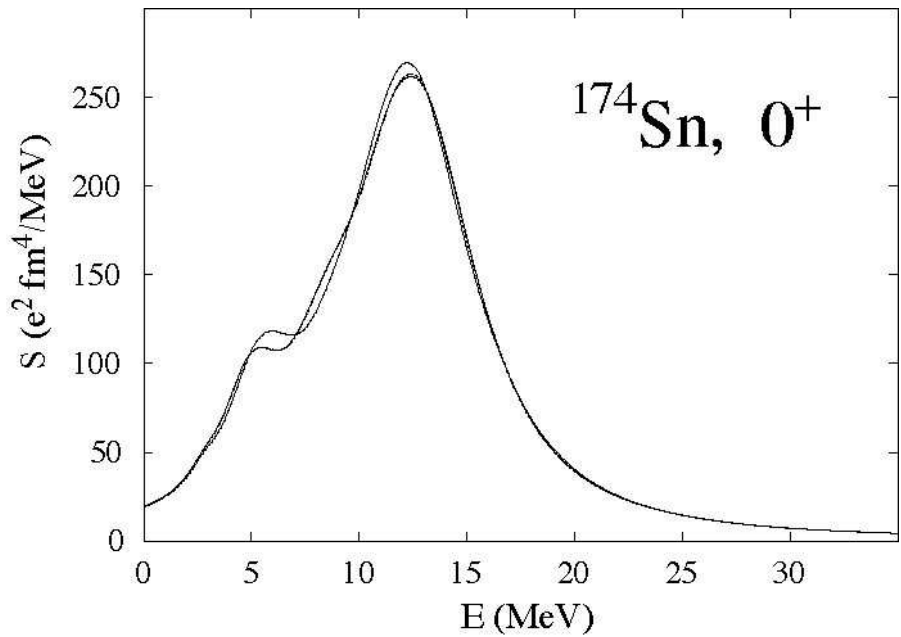}
\caption{\label{fig:0+_sn174_ecrit}  Isoscalar $0^+$ strength function
in $^{174}$Sn for (i) the  single-proton energy cutoff
$\varepsilon_{\rm crit}$=100 MeV and the neutron-quasiparticle occupation
cutoff $v^2_{\rm crit} = 10^{-8}$ (thin solid line); (ii)
$\varepsilon_{\rm crit}$=150 MeV and $v^2_{\rm crit} = 10^{-12}$  (dotted line);
 and (iii)
$\varepsilon_{\rm crit}$=200 MeV and $v^2_{\rm crit} = 10^{-16}$  (thick solid
line). Results corresponding to  (ii) and (iii) practically coincide.}
\end{figure}

Figure \ref{fig:0+_sn174_ecrit} shows the strength function
$S_J(E)$
for the isoscalar $0^+$ transition operator, cf.~\cite{har01},
\begin{equation}
\hat{F}_{00} = \frac{eZ}{A}\sum_{i=1}^A r_i^2.
\end{equation}
We have plotted three curves with successively more quasiparticle levels (from 246 proton
levels and 203 neutron levels to 341 proton levels and 374 neutron levels), with
cutoff parameters given in the figure caption.  The major structures in the
strength function are stable.  The error remaining after
to  the gentlest truncation
is extremely small.

\begin{figure}[t]
\includegraphics[width=15cm]{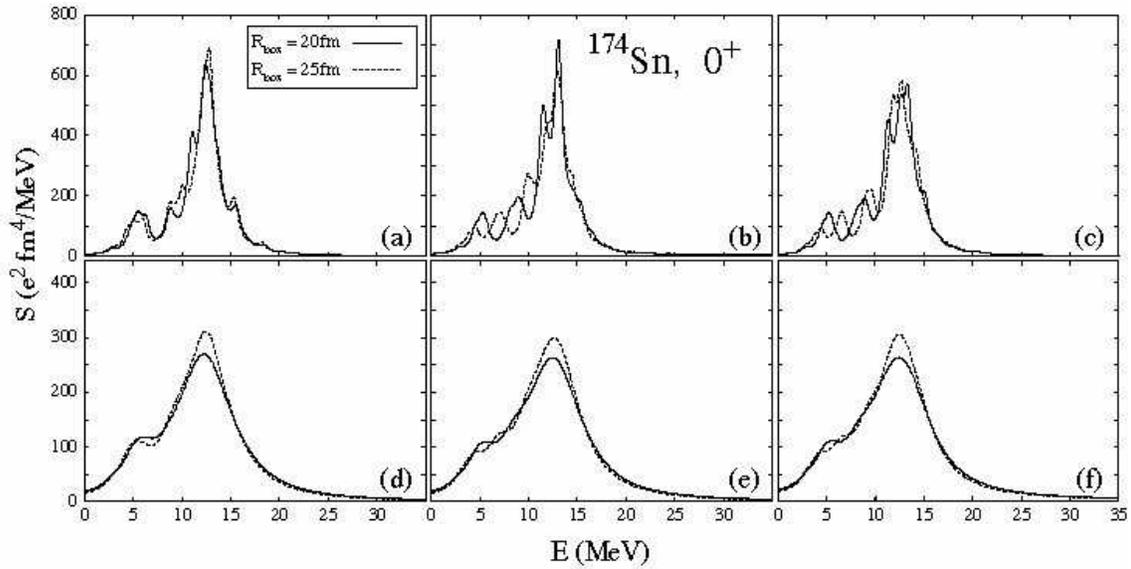}
\caption{\label{fig:0+_sn174_box} Isoscalar $0^+$ strength function
in $^{174}$Sn for the box radii: $R_{\rm box}$=20\,fm (solid line)
and  $R_{\rm box}$=25\,fm (dotted line).
In (a), (b), and (c) the smoothing-width parameter $\gamma$ is 0.5 MeV for all energies,
while in (d), (e), and (f) $\gamma(E)$ is given by Eq.~(\protect\ref{smwidth}).  We use the same three sets of cutoff conditions as in
Fig.~\protect\ref{fig:0+_sn174_ecrit}, namely
 (i) in parts  (a) and (d),
 (ii) in parts  (b) and (e),
and   (iii) in parts  (c) and (f).
 }
\end{figure}

The dependence of the strength function on the box size and
quasiparticle cutoff is shown in
Fig.~\ref{fig:0+_sn174_box}. The upper part of the Figure
(panels a-c) corresponds
to a constant smoothing width of $\gamma$=0.5 MeV. This relatively
small value is not sufficient to eliminate the finite-box effects
but it allows us to assess the stability of the QRPA solutions
as a function of $R_{\rm box}$.
The large structure corresponding to the giant monopole resonance (GMR)
is independent of box size no matter what the cutoff, but increasing the number of configurations magnifies the dependence on box size of local fluctuations in $S_J(E)$.
The lower part of the Figure (panels d-f) are smoothed more realistically, as in
Eq.~(\protect\ref{smwidth}). It is gratifying to see
that the resulting strength functions are practically identical,
i.e., the remaining dependence on $R_{\rm box}$ and the cutoff is very weak.
\begin{figure}[b]
\includegraphics[width=13cm]{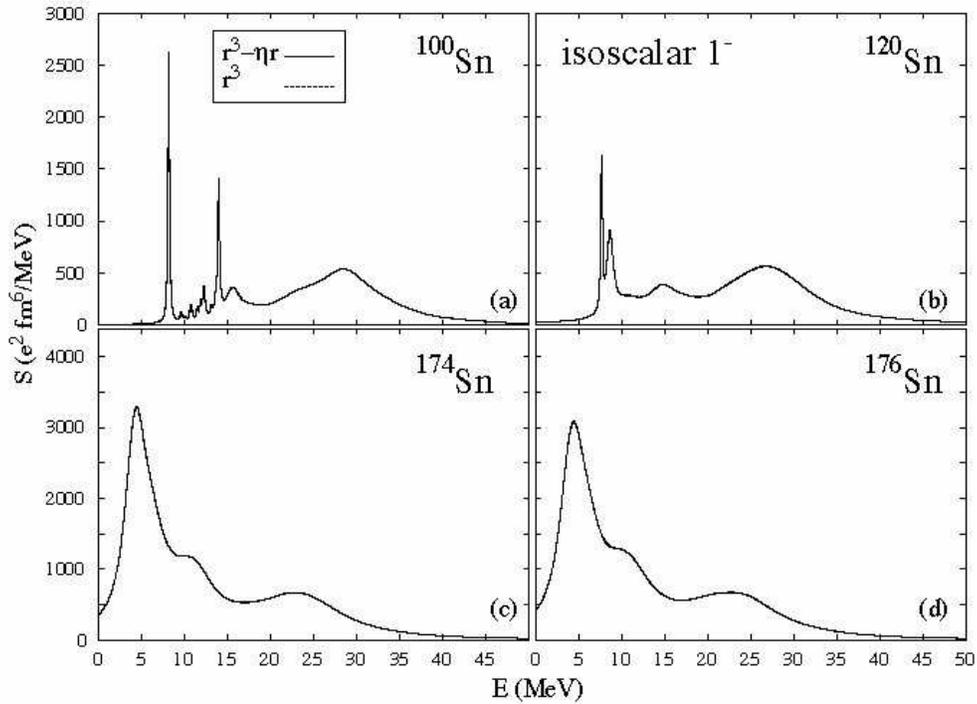}
\caption{\label{fig:1-_sn100_sn176}  Isoscalar $1^-$ strength function in
$^{100,120,174,176}$Sn
for the corrected dipole operator in Eq.\ (\protect\ref{dipolecorr}) (solid line)
and the uncorrected operator in Eq.\ (\protect\ref{dipole}) (dotted line). The cutoff
$\varepsilon_{\rm crit}$ is 140 MeV and $v^2_{\rm crit}$ is $3\times10^{-12}$.
The self consistency of our calculations makes the solid and dotted curves
coincide nearly exactly over the whole energy range.}
\end{figure}
\begin{figure}[b]
\includegraphics[width=15cm]{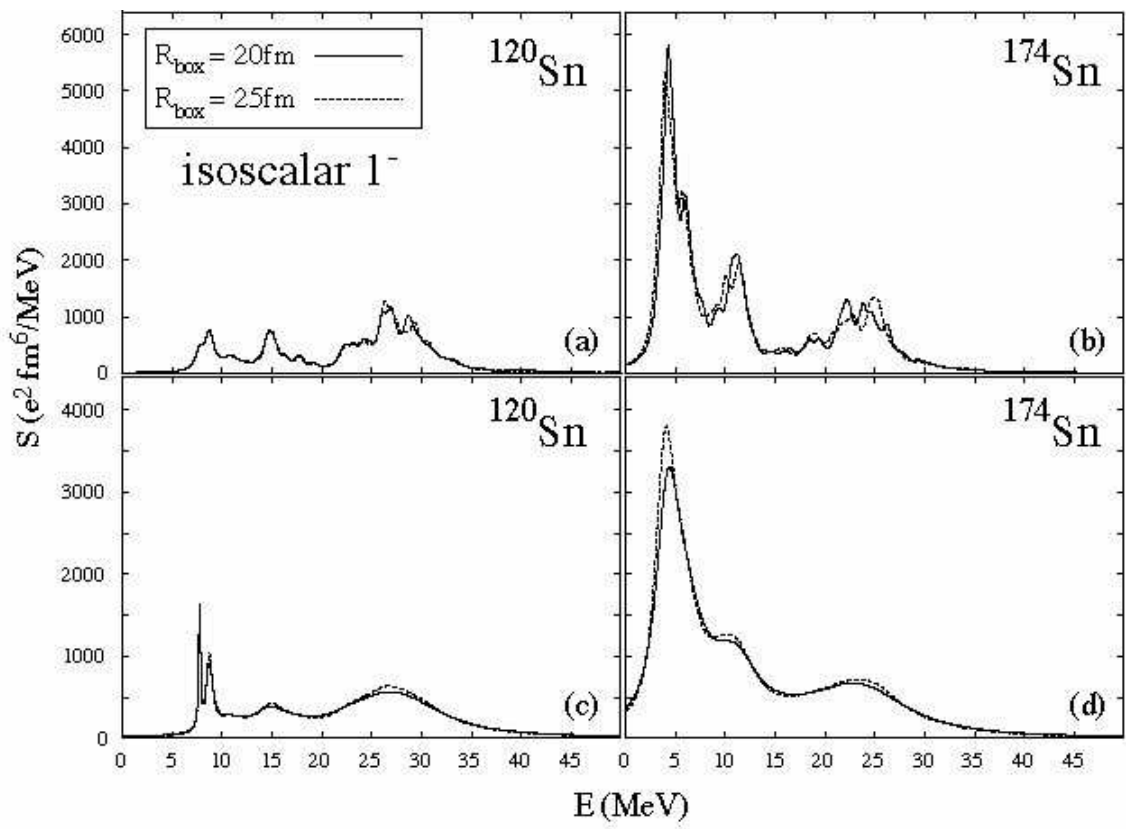}
\caption{\label{fig:0-_sn174_box} Isoscalar $1^-$ strength function
in $^{120}$Sn (left) and $^{174}$Sn (right)
for two box radii: $R_{\rm box}$=20\,fm (solid line)
and  $R_{\rm box}$=25\,fm (dotted line).
In (a) and (b) the smoothing-width parameter is constant ($\gamma$=0.5 MeV), while in (c) and (d) $\gamma(E)$ is given by Eq.~(\protect\ref{smwidth}).
 }
\end{figure}
\begin{figure}[b]
\includegraphics[width=14cm]{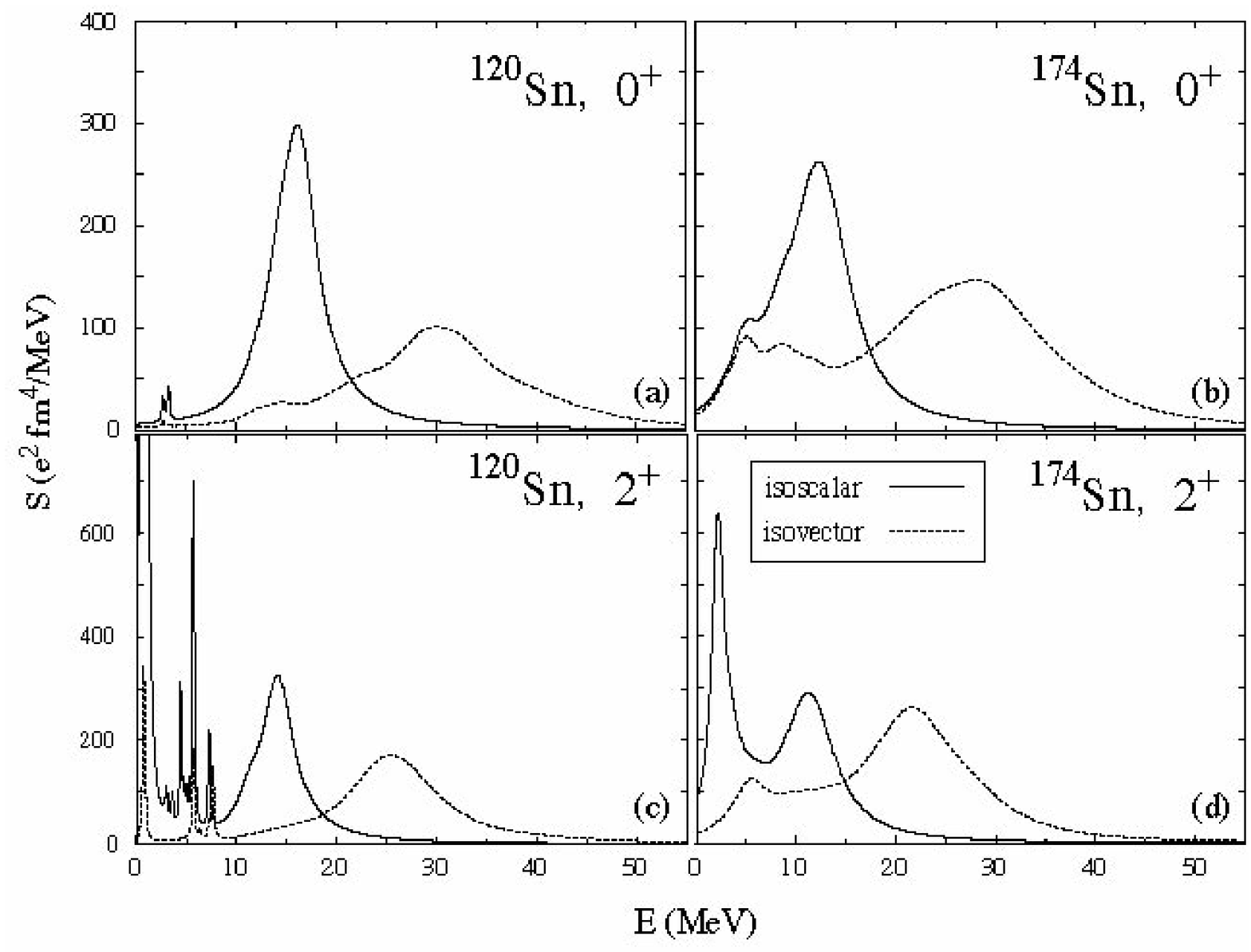}
\caption{\label{fig:is_iv_0+_2+}Isoscalar and isovector strength functions
for (a) the 0$^+$ channel of $^{120}$Sn, (b) the 0$^+$ channel of $^{174}$Sn,
(c) the 2$^+$ channel of $^{120}$Sn, and (d) the 2$^+$ channel of $^{174}$Sn.
The cutoff
$\epsilon_{\rm crit}$ is 150 MeV and $v^2_{\rm crit}$ is $10^{-12}$.
}
\end{figure}

The energy-weighted sum rule (EWSR) for the isoscalar $0^+$ mode \cite{har01} is given by
\begin{equation}
\sum_k  E_k|\langle k|\hat{F}_{00}|0\rangle|^2 =
2\frac{e^2\hbar^2}{m}\frac{Z^2}{A}\langle r^2 \rangle ,
\label{isoscalar_0+_sum_rule}
\end{equation}
where the expectation value is evaluated in the HFB ground
state.  This sum rule provides a stringent test of self consistency in the QRPA.
In $^{174}$Sn, the right-hand side of Eq.\
(\ref{isoscalar_0+_sum_rule}) is 35215 $e^2$~MeV~fm$^4$ and the left-hand side
34985$\pm 15$ $e^2$~MeV~fm$^4$ for all of the calculations of
Fig.~\ref{fig:0+_sn174_box};
the QRPA strength essentially exhausts the sum rule.
(The QRPA values of the EWSR in this paper are obtained by summing up to
$E_k=50$~MeV. )

\subsection{The isoscalar 1$^-$ mode}\label{oneminus}

The $1^-$ channel, home of the giant dipole resonance, the isoscalar squeezing
resonance, and as yet incompletely understood low-energy peaks in neutron-rich
nuclei (sometimes associated with skin excitations),
has a spurious isoscalar mode associated with center-of-mass motion that
can seriously compromise the low-energy spectrum if not handled with extreme
care.  We test the ability of our QRPA to do so in
$^{100}$Sn, $^{120}$Sn, $^{174}$Sn, and $^{176}$Sn.
(The nuclei $^{100}$Sn and $^{176}$Sn are the
two-proton and two-neutron drip-line systems predicted by the HFB calculation
with SkM$^{\ast}$. Neither nucleus has any static pairing,
i.e., $\Delta_{\rm n}$=$\Delta_{\rm p}$=0.) In the following calculations, we take
$\varepsilon_{\rm crit} = 140 $ MeV for the protons and
$v^2_{\rm crit}=9\times 10^{-12}$ for
the neutrons. As discussed above,  smoothed strength
functions are practically independent of small changes in the cutoff.
They are also independent of the cutoff in quasiparticle angular momentum
provided we include all states with $j$$\leq$15/2.

Figure \ref{fig:1-_sn100_sn176} shows the
predicted  isoscalar dipole strength
function for $^{100,120,174,176}$Sn.
%
%
For the transition operator, we use
\begin{equation}\label{dipole}
\hat{F}_{1M}=\frac{eZ}{A}\sum_{i=1}^A r_i^3Y_{1M}(\Omega_i)~,
\end{equation}
and the corrected operator
\begin{equation}\label{dipolecorr}
\hat{F}_{1M}^{\rm cor}=\frac{eZ}{A}\sum_{i=1}^A (r_i^3-\eta
r_i)Y_{1M}(\Omega_i), \ \
\eta = \frac{5}{3} \langle  r^2 \rangle ,
\end{equation}
to remove as completely as possible residual pieces of the spurious state from the physical states \cite{agr03}.  The fact that the strength functions
produced by these two operators --- displayed in Fig.~\ref{fig:1-_sn100_sn176}
 --- coincide so closely shows the extreme
accuracy of our QRPA solutions;
they are uncontaminated by spurious motion even
without the operator correction.  The spurious-state energies $E_{\rm spurious}$
are 0.964 MeV for $^{100}$Sn and 0.713 MeV for $^{120}$Sn, and the energies of
the first physical  excited states are 7.958 MeV for $^{100}$Sn and 7.729 MeV for
$^{120}$Sn.  In $^{174}$Sn ($^{176}$Sn), $E_{\rm spurious}$ is 0.319 MeV (0.349 MeV)
and the first physical state is at 3.485 MeV (2.710 MeV), lower than in
the more stable
isotopes.  Pairing correlations do not affect accuracy; the neutrons
in $^{120}$Sn  and $^{174}$Sn are paired, while those in $^{100}$Sn
and $^{176}$Sn are not.

We display the fine structure of the isoscalar $1^-$ strength functions in $^{120}$Sn and $^{174}$Sn in Fig.~\ref{fig:0-_sn174_box}, which also illustrates
the dependence of the results on $R_{\rm box}$. The dependence is consistent
with that of Fig.~\ref{fig:0+_sn174_box} for the isoscalar 0$^+$ strength;
the low-amplitude fluctuations in $S_J(E)$
that are unstable as a function of $R_{\rm box}$
disappear, and the smoothed strength function depends only weakly
on $R_{\rm box}$. In $^{120}$Sn, the two sharp peaks below
10 MeV correspond to discrete states while the broad maxima centered
around 15 MeV and 27 MeV are in the continuum, well above
neutron-emission threshold.  A similar three-peaked structure
emerges in $^{174}$Sn,
though most of the strength there is concentrated in the low-energy peak
at $E \approx 4$ MeV.  Fig.~\ref{fig:1-_sn100_sn176} shows (as we will discuss
in our forthcoming paper \cite{ter04}) that the appearance of
the low-energy isoscalar dipole strength is a real and dramatic feature
of neutron-rich dripline nuclei \cite{cat97,sag01}.
The EWSR for the isoscalar $1^-$ mode \cite{har01} is
\begin{equation}
\sum_k \sum_M E_k|\langle k|\hat{F}_{1M}^{\rm cor}|0\rangle|^2 =
\frac{3}{8\pi}\frac{e^2\hbar^2}{m}\frac{Z^2}{A}\left(
11\langle r^4\rangle - \frac{25}{3}\langle r^2\rangle^2\right).
\end{equation}
In $^{174}$Sn,
the right-hand side  is 403310 $e^2$~MeV~fm$^6$, while the
left-hand side  is 400200 $e^2$~MeV~fm$^6$. For
$^{176}$Sn, the corresponding numbers are 406576 $e^2$~MeV~fm$^6$ and
407100 $e^2$~MeV~fm$^6$.  This level of agreement is very good.

\subsection{The isovector 0$^+$  and isoscalar, isovector 2$^+$ modes}\label{twoplus}

Figure~\ref{fig:is_iv_0+_2+}
displays strength functions for the
0$^+$ and 2$^+$ channels in $^{120}$Sn and $^{174}$Sn.  (We discussed the
isoscalar $0^+$ mode above to illustrate the accuracy of our solutions, but
include it here as well for completeness.)
The calculations show the appearance of low-energy
$0^+$ strength --- both isovector and isoscalar --- and low-energy isovector
2$^+$ strength in $^{174}$Sn, though in none of these instances is the
phenomenon quite as dramatic as in the isoscalar $1^-$ channel.

The EWSR for the isoscalar 2$^+$
transition operator,
\begin{equation}
     \hat{F}_{2M}= e{\frac{Z}{A}} \sum_{i=1}^A r_i^2 Y_{2M}(\Omega_i),
\end{equation}
can be written as \cite{har01}
\begin{equation}\label{ewsr2}
\sum_k \sum_M E_k|\langle k|\hat{F}_{2M}|0\rangle|^2 =
\frac{25}{4\pi}\frac{e^2\hbar^2}{m}\frac{Z^2}{A}
\langle r^2\rangle.
\end{equation}
The sum rule is obeyed as well in the 2$^+$
isoscalar channel as in the $0^+$ and $1^-$ channels, the
only difference being that one needs to include quasiparticle states with $j > 15/2$ for
$^{174}$Sn.
For $^{120}$Sn ($^{174}$Sn) from Fig.\ \ref{fig:is_iv_0+_2+}, the EWSR is 37222 (34971) $e^2$\,MeV\,fm$^4$
while the QRPA value is 37030 (35010)
$e^2$\,MeV\,fm$^4$.

While on the topic of the sum rule, we
display 
in Table~\ref{tab:jmax_dependence_of_ewsr} the $j_{\rm max}$-dependence of
the EWSR
for several channels in $^{150}$Sn, with $R_{\rm box}= 25$ fm.
By taking $j_{\rm max}= 19/2$ we appear to obtain essentially 
the entire strength in all three
cases.
\begin{table}[h!]
\caption{\label{tab:jmax_dependence_of_ewsr}
The $j_{\rm max}$-dependence of isoscalar EWSR for $^{150}$Sn. $R_{\rm box}$ is
25 fm.
}
\parbox{10cm}{
\begin{ruledtabular}
\begin{tabular}{lcrr}
    $T$~$J^\pi$   & units  &$j_{\rm max}=19/2$ & $j_{\rm max}=25/2$ \\
\hline
IS 0$^+$ & $(e^2~{\rm MeV}~{\rm fm}^4)$ & 35731 & 35633 \\
IS 1$^-$ & $(e^2~{\rm MeV}~{\rm fm}^6)$ & 361686 & 353936 \\
IS 2$^+$ & $(e^2~{\rm MeV}~{\rm fm}^4)$ & 35542 & 35445 \\
\end{tabular}
\end{ruledtabular}
}
\end{table}

\section{\label{sec:conclusion} Conclusion}

In this work we have reported on the development and detailed testing
of a fully self-consistent Skyrme-QRPA framework that
employs the canonical HFB basis.  The method can be used to calculate strength
distributions in any spin-isospin channel and in any spherical even-even nucleus.
A good calculation requires a large single-quasiparticle space.
Our results show that our space is large enough in nuclei as heavy as the Sn isotopes.

We are
currently investigating the predictions of a range of Skyrme functionals across the
Ca, Ni, and Sn isotope chains.  The initial results presented here point to
increases in low-lying strength at the neutron drip line, particularly
in the isoscalar-dipole channel.  In a forthcoming paper \cite{ter04}
we will report
on the robustness of
these effects, on the physics underlying them, on their variation with atomic
mass and number, and on their implications for
the future of Skyrme functionals.

\begin{acknowledgments}
We gratefully acknowledge useful discussions with
Shalom Shlomo, Nils Paar, Dario Vretenar, and
 the Japanese
members of  the US-Japan cooperative project on
``Mean-Field Approach to Collective Excitations in Unstable 
Medium-Mass and Heavy Nuclei".
MB is grateful for the warm
hospitality at the Physics Division at ORNL and the Theory Division at 
GSI Darmstadt, where most of his contribution to this work was made.
This work was supported in part by the U.S. Department of Energy
under Contracts Nos.\  DE-FG02-97ER41019 (University of North Carolina at
Chapel Hill), DE-FG02-96ER40963 (University of Tennessee),
DE-AC05-00OR22725 with UT-Battelle, LLC (Oak Ridge National
Laboratory),  DE-FG05-87ER40361 (Joint Institute for Heavy Ion
Research), and
W-31-109-ENG-38 (Argonne National Laboratory);  
by the National Science Foundation Contract No.
0124053 (U.S.-Japan Cooperative Science Award);
by the Polish Committee for Scientific
Research (KBN); and  by the Foundation for Polish Science (FNP).
We used
the parallel computers of
The Center for Computational Sciences at Oak Ridge National Laboratory and
Information Technology Services at the University of North Carolina at Chapel Hill.

\end{acknowledgments}

\appendix

\appendix

\section{\label{app:QRPA_equation} QRPA equation}

The QRPA equations are the small-oscillations limit
of the time-dependent Hartree-Fock-Bogoliubov approximation, see,
e.g.,~\cite{rin80,war87}.
In the canonical basis the most general equations take the form
\begin{equation}
\sum_{L<L^\prime}
\left(
\begin{array}{cc}
A_{KK^\prime,LL^\prime} & B_{KK^\prime,LL^\prime} \\
-B^\ast_{KK^\prime,LL^\prime} & -A^\ast_{KK^\prime,LL^\prime}
\end{array}
\right)
\left(
\begin{array}{c}
{X}^k_{LL^\prime}\\
{Y}^k_{LL^\prime}
\end{array}
\right)
= E_k
\left(
\begin{array}{c}
{X}^k_{KK^\prime}\\
{Y}^k_{KK^\prime}
\end{array}
\right),
\ \ K<K^\prime ,
\label{QRPA_eq_with_m}
\end{equation}
\begin{eqnarray}
A_{KK^\prime,LL^\prime} &=&
E_{KL}\delta_{K^\prime L^\prime}
-E_{K^\prime L}\delta_{K L^\prime}
-E_{KL^\prime}\delta_{K^\prime L}
+E_{K^\prime L^\prime}\delta_{KL} \nonumber\\
&&-\bar{V}^{\rm ph}_{K \bar{L} \bar{K}^\prime L^\prime}u_{L^\prime}v_L u_K v_{K^\prime}
  +\bar{V}^{\rm ph}_{K^\prime \bar{L} \bar{K} L^\prime}u_{L^\prime}v_L u_{K^\prime} v_K \nonumber\\
&&+\bar{V}^{\rm ph}_{K \bar{L}^\prime \bar{K}^\prime L}u_{L}v_{L^\prime} u_K v_{K^\prime}
  -\bar{V}^{\rm ph}_{K^\prime \bar{L}^\prime \bar{K} L}u_L v_{L^\prime} u_{K^\prime} v_K \nonumber\\
&&-\bar{V}^{\rm pp}_{\bar{L} \bar{L}^\prime \bar{K}^\prime \bar{K}}v_{L}v_{L^\prime} v_{K^\prime} v_{K}
  -\bar{V}^{\rm pp}_{K K^\prime {L}^\prime L}u_K u_{K^\prime} u_{L} u_{L^\prime} \nonumber\\
&&-\bar{V}^{\rm 3p1h}_{\bar{L} \bar{L}^\prime {K} \bar{K}^\prime}v_{L}v_{L^\prime} u_{K} v_{K^\prime}
  +\bar{V}^{\rm 3p1h}_{\bar{L} \bar{L}^\prime {K}^\prime \bar{K}}v_L v_{L^\prime} u_{K^\prime} v_{K} \nonumber\\
&&-\bar{V}^{\rm 3p1h}_{{K} {K}^\prime \bar{L} {L}^\prime}u_{K}u_{K^\prime} u_{L^\prime} v_{L}
  +\bar{V}^{\rm 3p1h}_{{K} {K}^\prime \bar{L}^\prime {L}}u_{K} u_{K^\prime} u_{L} v_{L^\prime} \nonumber\\
&&-\bar{V}^{\rm 1p3h}_{\bar{L} {L}^\prime \bar{K}^\prime \bar{K}}u_{L^\prime}v_{L} v_{K^\prime} v_{K}
  +\bar{V}^{\rm 1p3h}_{\bar{L}^\prime {L} \bar{K}^\prime \bar{K}}u_{L} v_{L^\prime} v_{K^\prime} v_{K} \nonumber\\
&&-\bar{V}^{\rm 1p3h}_{{K} \bar{K}^\prime {L}^\prime {L}}u_{K}v_{K^\prime} u_{L} u_{L^\prime}
  +\bar{V}^{\rm 1p3h}_{{K}^\prime \bar{K} {L}^\prime {L}}u_{K^\prime} v_{K} u_{L} u_{L^\prime},
\label{A_matrix_with_m}
\end{eqnarray}
\begin{eqnarray}
B_{KK^\prime,LL^\prime} &=&
\bar{V}^{\rm ph}_{K^\prime L^\prime\bar{K} \bar{L}}u_{L^\prime}v_L u_{K^\prime} v_{K}
-\bar{V}^{\rm ph}_{K L^\prime \bar{K^\prime} \bar{L}}u_{L^\prime}v_{L} u_{K} v_{K^\prime}\nonumber\\
&&-\bar{V}^{\rm ph}_{K^\prime L \bar{K} \bar{L}^\prime}u_{L}v_{L^\prime} u_{K^\prime} v_{K}
+\bar{V}^{\rm ph}_{K L \bar{K}^\prime \bar{L}^\prime}u_{L}v_{L^\prime} u_{K} v_{K^\prime}\nonumber\\
&&+\bar{V}^{\rm pp}_{K^\prime {K} \bar{L} \bar{L}^\prime}v_{L}v_{L^\prime} u_{K} u_{K^\prime}
+\bar{V}^{\rm pp}_{L^\prime {L} \bar{K} \bar{K}^\prime}v_{K}v_{K^\prime} u_{L} u_{L^\prime}\nonumber\\
&&+\bar{V}^{\rm 3p1h}_{K^\prime {K} {L^\prime} \bar{L}}u_{L^\prime}v_{L} u_{K} u_{K^\prime}
-\bar{V}^{\rm 3p1h}_{K^\prime {K} {L} \bar{L}^\prime}u_{L}v_{L^\prime} u_{K} u_{K^\prime}\nonumber\\
&&+\bar{V}^{\rm 3p1h}_{L^\prime {L} {K^\prime} \bar{K}}u_{K^\prime}v_{K} u_{L} u_{L^\prime}
-\bar{V}^{\rm 3p1h}_{L^\prime {L} {K} \bar{K}^\prime}u_{K}v_{K^\prime} u_{L} u_{L^\prime}\nonumber\\
&&+\bar{V}^{\rm 1p3h}_{K^\prime \bar{K} \bar{L} \bar{L}^\prime}v_{L}v_{L^\prime} u_{K^\prime} v_{K}
-\bar{V}^{\rm 1p3h}_{K \bar{K}^\prime \bar{L} \bar{L}^\prime}v_{L}v_{L^\prime} u_{K} v_{K^\prime}\nonumber\\
&&+\bar{V}^{\rm 1p3h}_{L^\prime \bar{L} \bar{K} \bar{K}^\prime}v_{K}v_{K^\prime} u_{L^\prime} v_{L}
-\bar{V}^{\rm 1p3h}_{L \bar{L}^\prime \bar{K} \bar{K}^\prime}v_{K}v_{K^\prime} u_{L} v_{L^\prime},
\label{B_matrix_with_m}
\end{eqnarray}
\begin{equation}
 \bar{V}^{\rm ph}_{KL K^\prime L^\prime}
= \frac{ \delta^2 E[\rho,\kappa,\kappa^\ast] }{ \delta\rho_{K^\prime
K}\delta\rho_{L^\prime L} },
\label{ph_matrix_element_uncoupled}
\end{equation}
\begin{equation}
 \bar{V}^{\rm pp}_{K^\prime K L^\prime L}
= \frac{ \delta^2 E[\rho,\kappa,\kappa^\ast] }{ \delta\kappa^\ast_{K^\prime K}
 \delta\kappa_{L^\prime L} },
\label{pp_matrix_element_uncoupled}
\end{equation}
\begin{equation}
 \bar{V}^{\rm 3p1h}_{K^\prime K L^\prime L}
= \frac{ \delta^2 E[\rho,\kappa,\kappa^\ast] }{ \delta\kappa^\ast_{K^\prime K}
 \delta\rho_{L L^\prime} }
={\bar{V}^{\rm 1p3h\ \ast}}_{L L^\prime K^\prime K},
\label{3p1h_matrix_element_uncoupled}
\end{equation}
where $K$ and $L$ are single-particle indices for the canonical basis,
and the states are assumed to be ordered.
The symbol $\bar{K}$ refers to the conjugate partner of $K$,
$u_K$ and $v_K$ come from the BCS transformation associated with
the canonical basis, and the
$E_{KL}$ are the one-quasiparticle matrix elements of the HFB Hamiltonian
(cf. Eq. (4.14b) of Ref.~\cite{dob96}).
$X^k_{LL^\prime}$ and $Y^k_{LL^\prime}$ are the forward and backward
amplitudes of the QRPA solution $k$, and $E_k$ is the corresponding excitation energy.
$E[\rho,\kappa,\kappa^\ast]$ is the energy functional (see App.\
\ref{app:matrix_elements} for an explicit definition) and
$\rho$ and $\kappa$ are the density matrix and pairing tensor.
After taking the functional derivatives, we replace $\rho$ and $\kappa$
by their HFB solutions, in complete analogy with an ordinary Taylor-series
expansion.

To write the equations in coupled form, we introduce the notation
\begin{equation}
K \equiv (n_\mu l_\mu j_\mu m_\mu) \equiv (\mu m_\mu), \ \
L \equiv (\nu m_\nu), \ \
\end{equation}
where $(nljm)$ denote spherical quantum numbers.
Using (i) rotational, time-reversal, and parity symmetries of the HFB state,
(ii) the conjugate single-particle state\footnote{The
conjugate state in our HFB code is slightly different:
$|\overline{\mu m}\rangle = (-)^{j-m+l}
|\mu -m \rangle$.  This definition follows from Eq.~(3.24b) of Ref.~\cite{dob96} and
the single-particle wave function
$
\psi_{\mu m}({\bm r}) = R_{\mu}(r)\sum_{l_z\sigma}Y_{ll_z}(\Omega)
 \langle ll_z {\textstyle \frac{1}{2}} \sigma | j m \rangle |\sigma\rangle ,
$
where $R_\mu(r)$ and $|\sigma=\pm 1/2\rangle$ are real radial and spin wave
functions. Thus, we multiply all HFB $v_\mu$ by $(-)^{l}$ in the QRPA
calculations.}

\begin{equation}
|\overline{K}\rangle =
|\overline{\mu m_\mu}\rangle = (-)^{j_\mu-m_\mu}|\mu\: -m_\mu\rangle ,
\label{conjugate_QRPA}
\end{equation}
and (iii) the relations
\begin{eqnarray}
&&X^k_{KK^\prime} = \langle j_\mu m_\mu j_{\mu^\prime} m_{\mu^\prime}|J_k
M_k\rangle
\bar{X}^k_{[\mu\mu^\prime]J_k}
\times\left\{\begin{array}{cc}\sqrt{2}, &\mu=\mu^\prime,\\ 1,
&\mbox{otherwise},\end{array}\right.
\\
&&Y^k_{KK^\prime} = (-)^{j_\mu-m_\mu}(-)^{j_{\mu^\prime}-m_{\mu^\prime} }
\langle j_\mu -m_\mu j_{\mu^\prime} -m_{\mu^\prime}|J_k M_k\rangle
\bar{Y}^k_{[\mu\mu^\prime]J_k}
\times\left\{\begin{array}{cc}\sqrt{2}, &\mu=\mu^\prime,\\ 1,
&\mbox{otherwise},\end{array}\right.
\end{eqnarray}
with $J_k$ the angular momentum of the state $k$ and the factor $\sqrt{2}$
for convenience  \cite{row70},
one can rewrite the QRPA equation as

\begin{equation}
\sum_{\nu\leq\nu^\prime}
\left(
\begin{array}{cc}
A_{[\mu\mu^\prime]J_k,[\nu\nu^\prime]J_k}
&
B_{[\mu\mu^\prime]J_k,[\bar{\nu}\bar{\nu}^\prime]J_k}
\\
-B^\ast_{[\mu\mu^\prime]J_k,[\bar{\nu}\bar{\nu}^\prime]J_k}
&
-A^\ast_{[\mu\mu^\prime]J_k,[\nu\nu^\prime]J_k}
\end{array}
\right)
\left(
\begin{array}{c}
\bar{X}^k_{[\nu\nu^\prime]J_k}\\
\bar{Y}^k_{[\nu\nu^\prime]J_k}
\end{array}
\right)
= E_k
\left(
\begin{array}{c}
\bar{X}^k_{[\mu\mu^\prime]J_k}\\
\bar{Y}^k_{[\mu\mu^\prime]J_k}
\end{array}
\right),\ \ \mu \leq \mu^\prime ,
\label{QRPA_eq}
\end{equation}
\begin{eqnarray}
A_{[\mu\mu^\prime]J_k,[\nu\nu^\prime]J_k}
&=&
\frac{1}{\sqrt{1+\delta_{\mu\mu^\prime}}}
\frac{1}{\sqrt{1+\delta_{\nu\nu^\prime}}}
\left\{ E_{\mu\nu}\:\delta_{\mu^\prime\nu^\prime}
-E_{\mu^\prime\nu}\:\delta_{\mu\nu^\prime}(-)^{j_\mu+j_{\mu^\prime}-J_k}
\right.
\nonumber\\
&&
-E_{\mu\nu^\prime}\:\delta_{\mu^\prime\nu}(-)^{j_\mu+j_{\mu^\prime}-J_k}
+E_{\mu^\prime\nu^\prime}\:\delta_{\mu\nu}\nonumber\\
&& +G(\mu\mu^\prime\nu\nu^\prime;J_k)
(u_{\mu^\prime} u_\mu u_\nu u_{\nu^\prime}
+v_{\nu} v_{\nu^\prime} v_{\mu^\prime} v_{\mu}) \nonumber\\
&& +F(\mu\mu^\prime\nu\nu^\prime;J_k)
(u_{\mu} v_{\nu^\prime} u_\nu v_{\mu^\prime}
+u_{\mu^\prime} v_{\nu} u_{\nu^\prime} v_{\mu}) \nonumber\\
&&  -(-)^{ j_{\nu^\prime}+j_\nu-J_k }F(\mu\mu^\prime\nu^\prime\nu;J_k)
(u_{\mu} v_{\nu} u_{\nu^\prime} v_{\mu^\prime}
+u_{\mu^\prime} v_{\nu^\prime} u_{\nu} v_{\mu})
\nonumber\\
&&-H(\mu\mu'\nu\nu';J_k)( v_\nu v_{\nu'} u_\mu v_{\mu'} +u_{\mu'} v_\mu u_\nu u_{\nu'} )
\nonumber \\
&& +(-)^{j_\mu+j_{\mu'}-J_k} H(\mu'\mu\nu\nu';J_k)
(v_\nu v_{\nu'} u_{\mu'} v_\mu + u_\mu v_{\mu'} u_\nu u_{\nu'} )
\nonumber \\
&&-H^\ast(\nu\nu'\mu\mu';J_k)( u_\mu u_{\mu'} u_{\nu'} v_\nu
+ u_\nu v_{\nu'} v_{\mu'} v_\mu)
\nonumber \\
&&+(-)^{j_\nu+j_{\nu'}-J_k}H^\ast(\nu'\nu\mu\mu';J_k)
(u_\mu u_{\mu'} u_\nu v_{\nu'} + u_{\nu'} v_\nu v_{\mu'} v_\mu)
\left.\right\},
\label{A_spherical}
\end{eqnarray}
\begin{eqnarray}
B_{[\mu\mu^\prime]J_k,[\bar{\nu}\bar{\nu}^\prime]J_k}
&=&
\frac{1}{\sqrt{1+\delta_{\mu\mu^\prime}}}
\frac{1}{\sqrt{1+\delta_{\nu\nu^\prime}}}
\left\{
 -G(\mu\mu^\prime\nu\nu^\prime;J_k)
(u_{\mu^\prime} u_\mu v_\nu v_{\nu^\prime}
+u_{\nu} u_{\nu^\prime} v_{\mu^\prime} v_{\mu})\right. \nonumber\\
&& -(-)^{ j_\nu+j_{\nu^\prime}-J_k }F(\mu\mu^\prime\nu^\prime\nu;J_k)
(u_{\mu} u_{\nu} v_{\mu^\prime} v_{\nu^\prime}
+u_{\mu^\prime} u_{\nu^\prime} v_{\nu} v_{\mu}) \nonumber\\
&& +(-)^{ j_\nu+j_{\nu^\prime}+j_\mu+j_{\mu^\prime}
}F(\mu^\prime\mu\nu^\prime\nu;J_k)
(u_{\mu^\prime} u_{\nu} v_{\nu^\prime} v_{\mu}
+u_{\mu} u_{\nu^\prime} v_{\mu^\prime} v_{\nu})
\nonumber\\
&& +H(\mu\mu'\nu\nu';J_k)(v_\nu v_{\nu'} u_{\mu'} v_\mu + u_\mu v_{\mu'} u_\nu u_{\nu'})
\nonumber \\
&& -(-)^{j_\mu+j_{\mu'}-J_k} H(\mu'\mu\nu\nu';J_k)
(v_\nu v_{\nu'} u_\mu v_{\mu'} + u_{\mu'} v_\mu u_\nu u_{\nu'})
\nonumber\\
&&-H^\ast(\nu\nu'\mu\mu';J_k)
(v_\mu v_{\mu'} u_{\nu'} v_\nu + u_\nu v_{\nu'} u_\mu u_{\mu'})
\nonumber\\
&&+(-)^{j_\nu + j_{\nu'}-J_k}H^\ast(\nu'\nu\mu\mu';J_k)
(v_\mu v_{\mu'} u_\nu v_{\nu'} +u_{\nu'} v_\nu u_\mu u_{\mu'})
\left.\right\},
\label{B_spherical}
\end{eqnarray}
\begin{eqnarray}
G(\mu\mu^\prime\nu\nu^\prime;J_k)&=&\sum_{ m_\mu m_{\mu^\prime}m_\nu
m_{\nu^\prime} }
\langle j_\mu m_\mu j_{\mu^\prime} m_{\mu^\prime} | J_k M_k\rangle
\langle j_\nu m_\nu j_{\nu^\prime} m_{\nu^\prime} | J_k M_k\rangle
\bar{V}^{\rm pp}_{K K^\prime L L^\prime} \label{G_matrix_element} \nonumber\\
&\equiv& \langle [\mu\mu^\prime]J_k | \bar{V}^{\rm pp}
|[\nu\nu^\prime]J_k\rangle , \\
F(\mu\mu^\prime\nu\nu^\prime;J_k)&=&\sum_{ m_\mu m_{\mu^\prime}m_\nu
m_{\nu^\prime} }
\langle j_\mu m_\mu j_{\mu^\prime} m_{\mu^\prime} | J_k M_k\rangle
\langle j_\nu m_\nu j_{\nu^\prime} m_{\nu^\prime} | J_k M_k\rangle
\bar{V}^{\rm ph}_{K\bar{L^\prime}\bar{K^\prime}L} \nonumber\\
&=& \sum_{J^\prime} (-)^{j_{ \mu^\prime}+j_\nu +J^\prime }
 \left\{
 \begin{array}{ccc}
 j_\mu & j_{\mu^\prime} & J_k \\
 j_\nu & j_{\nu^\prime} & J^\prime
 \end{array}
 \right\}
 (2J^\prime +1) \langle [\mu\nu^\prime]J_k | \bar{V}^{\rm ph} | [\mu^\prime
\nu]J_k \rangle ,
\label{F_matrix_element} \\
H(\mu\mu'\nu\nu';J_k)
&=& \sum_{m_\mu m_{\mu'} m_\nu m_{\nu'}}
\langle j_\mu m_\mu j_{\mu'} m_{\mu'} | JM \rangle
\langle j_\nu m_\nu j_{\nu'} m_{\nu'} | JM \rangle
\bar{V}^{\rm 3p1h}_{\bar{L}\bar{L}' K \bar{K}'}  \nonumber\\
&=& \sum_{J^\prime} (-)^{j_{ \mu}+j_\nu + 1 + l_{\nu^\prime} -J_k - J^\prime }
 \left\{
 \begin{array}{ccc}
 j_\mu & j_{\mu^\prime} & J_k \\
 j_{\nu^\prime} & j_{\nu} & J^\prime
 \end{array}
 \right\}
 (2J^\prime +1) \langle [\mu\nu]J_k | \bar{V}^{\rm 3p1h} | [\mu^\prime
\nu^\prime]J_k \rangle. \nonumber\\
\label{H_matrix_element}
\end{eqnarray}
We have represented the second derivatives of the energy functional
$E[\rho,\kappa,\kappa^\ast]$ as unsymmetrized
matrix elements of effective interactions $\bar{V}^{\rm pp}$, $\bar{V}^{\rm
ph}$, and $\bar{V}^{\rm 3p1h}$. These effective interactions are given
in App.\ \ref{app:matrix_elements}.
Although the ``matrix elements'' are unsymmetrized, the underlying
two-quasiparticle states are of course antisymmetric.  As a consequence, $A_{[\mu\mu^\prime]J_k,[\nu\nu^\prime]J_k}=
B_{[\mu\mu^\prime]J_k,[ \bar{\nu} \bar{\nu}^\prime ]J_k}=0$
if $J_k$ is odd and either $\mu=\mu^\prime$ or $\nu=\nu^\prime$.

The nuclear energy functional $E[\rho,\kappa,\kappa^\ast]$
is usually separated into
particle-hole (ph) and pairing pieces (again, see App.~B for
explicit expressions).
If the pairing functional, which we will call $E_{\rm pair}[\rho,\kappa,\kappa^\ast]$, depends on $\rho$
then the derivatives of $E_{\rm pair}[\rho,\kappa,\kappa^\ast]$ with respect to $\rho_{KK'}$
are called pairing-rearrangement terms \cite{war87}.
In the QRPA, two kinds of pairing-rearrangement terms can arise in general.
One has particle-hole character and is included in $\bar{V}^{\rm ph}_{KLK'L'}$;
the other affects 3-particle-1-hole (3p1h) and 1-particle-3-hole (1p3h)
configurations and is
represented by $\bar{V}^{\rm 3p1h}_{K'KL'L}$ and $\bar{V}^{\rm 1p3h}_{K'KL'L}$.
If the $\rho$-dependence of $E_{\rm pair}[\rho,\kappa,\kappa^\ast]$ is linear,
then the ph-type pairing-rearrangement term does not appear.
Furthermore, the 3p1h and 1p3h pairing-rearrangement terms arise only for
$J^{\pi}=0^+$ modes
if the HFB state has $J=0$.
Most existing work uses a pairing functional that is linear in $\rho$,
and so needs no
pairing-rearrangement terms in $J^{\pi} \neq 0^+$ channels.

\section{\label{app:matrix_elements} Interaction matrix elements (second
functional derivatives)}
\subsection{Representation of second derivatives as matrix element of
effective interactions}

In this appendix, we discuss interaction matrix elements coming from
$E[\rho,\kappa,\kappa^\ast]$,
which we take to contain separate Skyrme (i.e.\ strong-force,
$\kappa$-independent), Coulomb, and pairing energy functionals:
\begin{equation}
E[\rho,\kappa,\kappa^\ast]
= E_{\rm Skyrme}[\rho] + E_{\rm Coul}[\rho_{\rm p}] +E_{\rm pair}[\rho,\kappa,\kappa^\ast],
\end{equation}
where $\rho_{\rm p}$ is the proton density matrix.
The most general Skyrme energy functional in common use is given by
\begin{eqnarray}
\label{e:skyrme}
E_{\rm Skyrme}[\rho]
&=& \sum_{t=0,1} \int d^3r
    \bigg\{  C_t^{\rho}[\rho_{00}]\, \rho^2_{t0} (\svec{r})
           + C_t^{\Delta\rho} \rho_{t0}(\svec{r}) \Delta \rho_{t0}(\svec{r})
          + C_t^{\tau} \left[  \rho_{t0}(\svec{r}) \, \tau_{t0}(\svec{r})
                              -\svec{j}^2_{t0}(\svec{r})
                        \right]
    \nonumber \\
& &        + C_t^{s}[\rho_{00}]\, \svec{s}^2_{t0}(\svec{r})
           + C_t^{\Delta s}\svec{s}_{t0}(\svec{r})\cdot\Delta\svec{s}_{t0}(\svec{r})
           + C_t^{T} \left[ \svec{s}_{00}(\svec{r}) \cdot \svec{T}_{t0}(\svec{r})
                          -\tensor{J}{}^2_{t0}(\svec{r} )
                     \right]
    \nonumber \\
& &        + C_t^{\nabla J}
                     \left[ \rho_{t0}(\svec{r})\svec\nabla\cdot\svec{J}_{t0}(\svec{r})
                          +\svec{s}_{t0} (\svec{r}) \cdot \svec\nabla
                           \times \svec{j}_{t0}(\svec{r})
                     \right]
    \bigg\} .
\end{eqnarray}
(See, e.g., \cite{dob95,ben03,per04} and references therein for a general
discussion.)
All densities are labeled by isospin indices $tt_z$, where $t$
takes values zero and one and $t_z$ is always equal to 0.
A more general theory could violate isospin at the single-quasiparticle level,
leading to additional densities $\rho_{1\pm 1}$ \cite{per04}.
We do not consider such densities here.
The $C_t^{i}$ are the coupling constants for the effective interaction.
As usual, two of them are chosen to be density dependent:
\begin{eqnarray}
C_t^\rho [\rho_{00}]
& = & A_t^\rho + B_t^\rho \, \rho_{00}^\alpha(\svec{r}) ,
      \nonumber \\
C_t^s [\rho_{00}]
& = & A_t^s + B_t^s \, \rho_{00}^\alpha(\svec{r}) .
\end{eqnarray}
Here $\rho_{t0}$, $\svec{s}_{t0}$, $\tau_{t0}$, $\svec{T}_{t0}$, $\svec{j}_{t0}$,
$\tensor{J}_{t0}$, and $\svec{J}_{t0}$ are local densities and currents, which
are derived from the general density matrices for protons and neutrons
\begin{eqnarray}
\rho_{00} (\svec{r}\sigma,\svec{r}'\sigma')
& = &   \rho_{\text{n}} (\svec{r}\sigma,\svec{r}'\sigma')
      + \rho_{\text{p}} (\svec{r}\sigma,\svec{r}'\sigma'),
     \nonumber \\
\rho_{10} (\svec{r}\sigma,\svec{r}'\sigma')
& = &   \rho_{\text{n}} (\svec{r}\sigma,\svec{r}'\sigma')
      - \rho_{\text{p}} (\svec{r}\sigma,\svec{r}'\sigma'),
\end{eqnarray}
where
\begin{eqnarray}
\rho_{\text n}(\svec{r}\sigma,\svec{r}^\prime\sigma^\prime)
&=& \sum_{KK^\prime,\text{neutron}} \psi^\ast_{K^\prime}(\svec{r}^\prime\sigma^\prime)
 \psi_{K}(\svec{r}\sigma)\rho_{K K^\prime} , \nonumber \\
\rho_{\text p}(\svec{r}\sigma,\svec{r}^\prime\sigma^\prime)
&=& \sum_{KK^\prime,\text{proton}} \psi^\ast_{K^\prime}(\svec{r}^\prime\sigma^\prime)
 \psi_{K}(\svec{r}\sigma)\rho_{K K^\prime} ,
\end{eqnarray}
and $\sigma = \pm \frac{1}{2}$ labels the spin components so that, e.g.,
$\psi_K(\svec{r}\sigma)$ is a spin component of the single-particle wave
function associated with the state
$K$.
Defining
\begin{eqnarray}
\rho_{t0} (\svec{r},\svec{r}')
& = & \sum_{\sigma = \pm} \rho_{t0} (\svec{r}\sigma,\svec{r}'\sigma) ,
      \nonumber \\
\svec{s}_{t0} (\svec{r},\svec{r}')
& = & \sum_{\sigma,\sigma' = \pm} \rho_{t0} (\svec{r}\sigma,\svec{r}'\sigma') \,
      \sigmamat
,
\end{eqnarray}
where $\sigmamat = \langle \sigma^\prime | \svec{\sigma} | \sigma \rangle $ is a matrix element of the vector of
Pauli spin matrices, we write the local densities and currents as
\begin{eqnarray}
\rho_{t0}(\svec{r})
& = & \rho_{t0} (\svec{r},\svec{r}),
     \nonumber \\
\svec{s}_{t0}(\svec{r})
& = & \svec{s}_{t0} (\svec{r},\svec{r}),
      \nonumber \\
\tau_{t0}(\svec{r})
& = & \svec\nabla \cdot \svec\nabla' \rho_{t0}(\svec{r},\svec{r}')
      |_{\svec{r}=\svec{r}'} ,
      \nonumber \\
\svec{T}_{t0}(\svec{r})
& = & \svec\nabla \cdot \svec\nabla' \svec{s}_{t0}(\svec{r},\svec{r}')
      |_{\svec{r}=\svec{r}'},
      \nonumber \\
\svec{j}_{t0} (\svec{r})
& = & - \tfrac{i}{2} (\svec\nabla-\svec\nabla') \, \rho_{t0}(\svec{r},\svec{r}')
      |_{\svec{r}=\svec{r}'},
      \nonumber \\
J_{t0,ij}(\svec{r})
& = & -\tfrac{i}{2}
      (\svec\nabla - \svec\nabla')_i \, s_{t0,j} (\svec{r},\svec{r}')
      |_{\svec{r}=\svec{r}'} ,
      \nonumber \\
\stackrel{\leftrightarrow}{J}{\!}^2_{t0}(\svec{r})
&=& \sum_{ij=xyz} J^2_{t0,ij},
\nonumber \\
\svec{J}_{t0}(\svec{r})
& = & -\tfrac{i}{2} (\svec\nabla-\svec\nabla') \times \svec{s}_{t0} (\svec{r},\svec{r}' )
      |_{ \svec{r}=\svec{r}' } .
\end{eqnarray}

The Coulomb energy functional is given by
\begin{equation}
\label{e:coulomb}
E_{\text{Coul}} [\rho_{\rm p}]
=  \frac{e^2}{2} \iint \! d^3r \, d^3 r'
   \frac{\rho_{\rm p}(\svec{r}) \rho_{\rm p}(\svec{r}^\prime)}
        {|\svec{r}-\svec{r}'|}
  -\frac{3}{4} e^2 \left(\frac{3}{\pi}\right)^{\frac{1}{3}}
   \int \! d^3 r \; \rho_{\text{p}}^{4/3} (\svec{r})
,
\end{equation}
where we make the usual Slater approximation \cite{sla51} for  the exchange term.

For the  pairing functional we take the quite general form
\begin{equation}
E_{\text{pair}}[\rho,\kappa,\kappa^\ast]
= E_{\text{pair}}[\rho,\tilde{\rho},\tilde{\rho}^\ast]
=  \int \! d^3 r \;
  C^{\tilde\rho}[\rho_{00}(\svec{r})]
  \sum_{\tau={\rm p,n}} |\tilde{\rho}_\tau(\svec{r})|^2 ,
\label{pairing_energy_functional}
\end{equation}
where the density-dependent pairing coupling constant
 $C^{\tilde\rho}[\rho_{00}(\svec{r})]$ is an arbitrary function of
$\rho_{00}(\svec{r})$.
The quantity
$\tilde{\rho}_\tau(\svec{r})$
is defined as \cite{dob84}
\begin{eqnarray}
\tilde{\rho}_\tau(\svec{r})
& = &-i \sum_{\sigma\sigma'=\pm}
          \kappa_{\tau}(\svec{r}\sigma,\svec{r}\sigma')
          \sigma^y_{\sigma\sigma'} ,\ \tau={\rm proton \ or \ neutron},
\end{eqnarray}
with
\begin{eqnarray}
\kappa_{\text n}(\svec{r}\sigma,\svec{r}^\prime\sigma^\prime)
&=& \sum_{KK^\prime,\text{neutron}}
\psi_{K^\prime}(\svec{r}^\prime\sigma^\prime)
\psi_{K}(\svec{r}\sigma)\kappa_{KK^\prime} , \nonumber \\
\kappa_{\text p}(\svec{r}\sigma,\svec{r}^\prime\sigma^\prime)
&=& \sum_{KK^\prime,\text{proton}}
\psi_{K^\prime}(\svec{r}^\prime\sigma^\prime)
 \psi_{K}(\svec{r}\sigma)\kappa_{KK^\prime},
\end{eqnarray}
being the standard pairing tensor in the coordinate representation.

The second derivatives of the energy functional in Eq.\
(\ref{ph_matrix_element_uncoupled}), as the equation indicates and we've
already noted, can be written
as \emph{unsymmetrized} matrix elements $\bar{V}^{\rm ph}_{KL K'L'}$
of an effective interaction between uncoupled pairs of single-particle
states.  The particle-hole matrix elements take the form
\begin{equation}
\bar{V}^{\rm ph}_{KL K' L'}
= \langle KL|
    \bar{V}^{\text{eff}}_{\text{Skyrme}}
                 + \bar{V}^{\text{eff}}_{\text{Coul}}
                 + \bar{V}^{\rm eff\ ph}_{\rm pair}  |K' L'\rangle.
\label{decomp}
\end{equation}
The last term contains the pairing rearrangement discussed at the end of the
previous appendix.

The effective Skyrme interaction in Eq.\ (\ref{decomp}) is given by
\begin{eqnarray}
\label{Skyrme_operator}
\bar{V}^{\text{eff}}_{\text{Skyrme}}
& = &   ( a_0
         +b_0 \, \svec\sigma \cdot \svec\sigma'
         +c_0 \, \ivec\tau   \cdot \ivec\tau'
         +d_0 \, \svec\sigma \cdot \svec\sigma' \, \ivec\tau \cdot \ivec\tau'
        ) \, \delta(\svec{r}-\svec{r}')
      \nonumber\\
&   & + (  a_1
          +b_1 \, \svec\sigma \cdot \svec\sigma'
          +c_1 \, \ivec\tau   \cdot \ivec\tau'
          +d_1 \, \svec\sigma \cdot \svec\sigma' \, \ivec\tau \cdot \ivec\tau'
      ) \, ( {\kdag}^2 \, \delta(\svec{r}-\svec{r}')
              + \delta(\svec{r}-\svec{r}')\svec{k}^2 )
      \nonumber\\
&   & + (  a_2
          +b_2 \, \svec\sigma \cdot \svec\sigma'
          +c_2 \, \ivec\tau   \cdot \ivec\tau'
          +d_2 \, \svec\sigma \cdot \svec\sigma' \, \ivec\tau \cdot \ivec\tau'
        ) \,  \kdag \cdot \delta(\svec{r}-\svec{r}') \, \svec{k}
      \nonumber\\
&   & + (  a_3
          +b_3 \, \svec\sigma \cdot \svec\sigma'
          +c_3 \, \ivec\tau   \cdot \ivec\tau'
          +d_3 \, \svec\sigma \cdot \svec\sigma' \, \ivec\tau \cdot \ivec\tau'
        ) \, \rho_{00}^\alpha(\svec{r}) \, \delta(\svec{r}-\svec{r}')
      \nonumber\\
&   & + \big[ e_3 \, \rho_{10}(\svec{r}) \, (\tau^{(0)}+\tau^{\prime(0)})
             +g_3 \, \svec{s}_{00} (\svec{r}) \cdot (\svec\sigma + \svec\sigma')
      \nonumber\\
&   & \qquad
             +m_3 \, \svec{s}_{10} (\svec{r}) \cdot
                     (\svec\sigma \tau^{(0)} + \svec\sigma' \tau^{\prime(0)})
        \big] \, \rho_{00}^{\alpha-1}(\svec{r}) \, \delta(\svec{r}-\svec{r}')
      \nonumber\\
&   & + \big[ f_3 \, \rho^2_{10}(\svec{r})
             +h_3 \, \svec{s}^2_{00}(\svec{r})
             +n_3 \, \svec{s}^2_{10}(\svec{r})
        \big] \, \rho_{00}^{\alpha-2}(\svec{r}) \, \delta(\svec{r}-\svec{r}')
      \nonumber\\
&   & +(   a_4
          +c_4 \, \ivec\tau \cdot \ivec\tau') \,
                  (\svec\sigma + \svec\sigma') \cdot
                  \kdag \times \delta(\svec{r}-\svec{r}') \, \svec{k}
.
\label{operator_V_skyrme}
\end{eqnarray}
$\ivec\tau=(\tau^{(\pm 1)},\tau^{(0)})$ is the vector of
Pauli matrices in isospin space and
\begin{eqnarray}
\svec{k}
& = & - \tfrac{i}{2} (\svec\nabla-\svec\nabla')
      \quad \text{acting to the right,}
      \nonumber\\
\kdag
& = & \phantom{-} \tfrac{i}{2} (\svec\nabla-\svec\nabla')
      \quad \text{acting to the left}.
\end{eqnarray}
The coefficients in Eq.~(\ref{Skyrme_operator}) are defined in
Tables~\ref{tab:coefficients_abc} and~\ref{tab:coefficients_efg}.
Equation (\ref{operator_V_skyrme}) contains the usual
Skyrme-interaction operators,
but, the energy functional (\ref{e:skyrme}) does not necessarily correspond
to a real (density-dependent) two-body Skyrme interaction because the
matrix elements
are not antisymmetrized.
Compared to the case usually discussed in the literature, the more general
functional relaxes relations that would otherwise restrict the spin-isospin structure of
the effective interaction
in Eq.~(\ref{Skyrme_operator}); see, e.g., \cite{ben02}
for a discussion of the increased freedom.

The densities and currents that appear in Eq.\ (\ref{Skyrme_operator})
come mostly from rearrangement terms and take
the values given by the HFB ground state.
The isoscalar and isovector spin densities
$\svec{s}_{t0} (\svec{r})$ vanish when the
HFB ground state is time-reversal invariant or spherical as assumed here.
The terms containing them will therefore not appear in the expressions for the matrix elements
of the effective interaction for such states given below.

\begin{table}
\caption{\label{tab:coefficients_abc} Definitions of $a_i, b_i, c_i, d_i$
$(i=0,\cdots,3)$, $a_4$, and $c_4$ in Eq.~(\ref{Skyrme_operator}).
}
\begin{ruledtabular}
\begin{tabular}{ccccc}
$i$ & $a_i$ & $b_i$ & $c_i$ & $d_i$ \\
\hline
0 & $\scriptst 2A_0^{\rho}$ & $\scriptst 2 A_0^{s}$ &
    $\scriptst 2A_1^{\rho}$ & $\scriptst 2 A_1^{s}$ \\
1 & $\scriptst \frac{1}{2}(C_0^{\tau}- 4C_0^{\Delta\rho})$ &
    $\scriptst \frac{1}{2}(C_0^{T}   - 4C_0^{\Delta s})$   &
    $\scriptst \frac{1}{2}(C_1^{\tau}- 4C_1^{\Delta\rho})$ &
    $\scriptst \frac{1}{2}(C_1^{T}   - 4C_1^{\Delta s})$ \\
2 & $\scriptst 3C_0^{\tau} + 4C_0^{\Delta\rho}$ &
    $\scriptst 3C_0^{T}    + 4C_0^{\Delta s}$   &
    $\scriptst 3C_1^{\tau} + 4C_1^{\Delta\rho}$ &
    $\scriptst 3C_1^{T}    + 4C_1^{\Delta s}$ \\
3 & $\scriptst  B_0^{\rho}(\alpha+2)(\alpha+1)$ &
    $\scriptst 2B_0^{s}$    &
    $\scriptst 2B_1^{\rho}$ &
    $\scriptst 2B_1^{s}$    \\
4 & $\scriptst -2iC_0^{\nabla J}$ & &
    $\scriptst -2iC_1^{\nabla J}$ &  \\
\end{tabular}
\end{ruledtabular}
\end{table}

\begin{table}
\caption{\label{tab:coefficients_efg} Definitions
of the coefficients appearing in the rearrangement terms
in Eq.~(\ref{Skyrme_operator}).
}
\begin{ruledtabular}
\begin{tabular}{ccccccc}
$i$ & $e_i$ & $f_i$ & $g_i$ & $h_i$ & $m_i$ & $n_i$ \\
\hline
3 & $\scriptst 2\alpha B_1^{\rho}$ &
    $\scriptst \alpha(\alpha-1) B_1^{\rho}$ &
    $\scriptst 2\alpha B_0^{s}$ &
    $\scriptst \alpha(\alpha-1) B_0^{s}$ &
    $\scriptst 2\alpha B_1^{s}$ &
    $\scriptst \alpha(\alpha-1) B_1^{s}$  \\
\end{tabular}
\end{ruledtabular}
\end{table}

The effective Coulomb interaction in Eq.\ (\ref{decomp}), acting between
protons, is given by
\begin{equation}
\bar{V}^{\text{eff}}_{\rm Coul}
=  \frac{e^2}{|\svec{r}-\svec{r}'|}
  -\frac{e^2}{3} \left(\frac{3}{\pi}\right)^{\frac{1}{3}}
   \rho_{\text{p}}^{-2/3} (\svec{r}) \, \delta(\svec{r}-\svec{r}')
.
\end{equation}
Finally, the ph-type pairing-rearrangement terms in Eq.\ (\ref{decomp}) come
from an effective interaction
\begin{equation}
\bar{V}^{\rm eff \ ph}_{\rm pair} =
\frac{d^2 C^{\tilde\rho}[\rho_{00}(\svec{r})]}{d \rho_{00}^2(\svec{r})}
\sum_{\tau={\rm p,n}}|\tilde{\rho}_\tau(\svec{r})|^2\, \delta(\svec{r}'-\svec{r}).
\end{equation}

The second derivatives with respect to $\kappa,\kappa^\ast$ also can be
written as unsymmetrized matrix elements of effective interactions, this time
in the particle-particle channel.  The particle-particle effective interaction entering the matrix elements
\begin{equation}
\bar{V}^{\rm pp}_{KK^\prime LL^\prime}
= \langle KK^\prime|
  \bar{V}^{\text{eff\ pp}}_{\text{pair}}   |LL^\prime\rangle
\end{equation}
is obtained from Eq.~(\ref{pairing_energy_functional})
through Eq.~(\ref{pp_matrix_element_uncoupled}) as
\begin{equation}
\bar{V}^{\text{eff\ pp}}_{\text{pair}}= C^{\tilde\rho}[\rho_{00}(\svec{r})]
  \left(  3
         - \svec\sigma \cdot \svec\sigma'
         - \ivec\tau   \cdot \ivec\tau'
         - \svec\sigma \cdot \svec\sigma' \, \ivec\tau   \cdot \ivec\tau'
  \right) \, \delta(\svec{r}-\svec{r}')
.
\end{equation}
In the numerical calculations of this paper, we use a volume pairing-energy functional,
i.e.,
\begin{equation}
C^{\tilde\rho} = \frac{1}{2}V_0 = {\rm const.}
\label{volpair}
\end{equation}

Last of all are the mixed functional derivatives involving both $\rho$ and
$\kappa$ (or $\tilde{\rho}$)
 in Eq.\ (\ref{3p1h_matrix_element_uncoupled}).  They also can be
written as the unsymmetrized matrix elements of an effective interaction:
\begin{equation}
\bar{V}^{\rm 3p1h}_{K'KL'L} = \langle L'K'| \bar{V}^{\rm eff\ 3p1h}_{\rm pair
} |LT(K)\rangle , \label{v_3p1h_matrix_element}
\end{equation}
%
where $T(K)$ denotes the time-reversed state of $K$, and the
3p1h effective interaction itself is
\begin{equation}
\bar{V}^{\rm eff\ 3p1h}_{\rm pair}
 = \frac{dC^{\tilde\rho}[\rho_{00}(\svec{r})]}{d\rho_{00}(\svec{r})}
\left[ \tilde{\rho}_{\rm p} (\svec{r})\left(1-\tau_z^\prime\right)
+ \tilde{\rho}_{\rm n}
(\svec{r})\left(1+\tau_z^\prime\right)\right]
\delta(\svec{r}'-\svec{r}) ,
\end{equation}
where $\tau'_z$ acts on the single-particle states $K'$ and $T(K)$ in
Eq.~(\ref{v_3p1h_matrix_element}), and the eigenvalues 1 and $-1$ are
assigned to the neutron and proton, respectively.

\subsection{Calculation of matrix elements}

To calculate the coupled matrix elements in Eqs.\ (\ref{G_matrix_element})--(\ref{H_matrix_element}),
we use an intermediate LS scheme:
\begin{eqnarray}
 \lefteqn{ \langle [\mu\mu^\prime]J_k|\bar{V}|[\nu\nu^\prime]J_k\rangle }
 \nonumber\\
&=& \sum_{LL^\prime S} \hat{j}_\mu \hat{j}_{\mu^\prime} \hat{j}_\nu
\hat{j}_{\nu^\prime}
 \hat{L} \hat{L^\prime} \hat{S}^2
\left\{
\begin{array}{ccc}
l_\mu & l_{\mu^\prime} & L \\
1/2 & 1/2 & S \\
j_\mu & j_{\mu^\prime} & J_k
\end{array}
\right\}
\left\{
\begin{array}{ccc}
l_\nu & l_{\nu^\prime} & L^\prime \\
1/2 & 1/2 & S \\
j_\nu & j_{\nu^\prime} & J_k
\end{array}
\right\}\nonumber \\
&&\times \langle (l_\mu l_{\mu^\prime})LS;J_k|\bar{V}|(l_\nu
l_{\nu^\prime})L^\prime S;J_k\rangle,
\end{eqnarray}
\begin{equation}
\hat{j}_\mu \equiv \sqrt{2j_\mu + 1}.
\end{equation}
Eq.~(\ref{Skyrme_operator}) gives

\noindent
i) proton-proton or neutron-neutron matrix elements:
\begin{eqnarray}
\lefteqn{ \langle (l_\mu l_{\mu^\prime})LS;J_k|
\bar{V}^{\text{eff}}_{\text{Skyrme}}
|(l_\nu l_{\nu^\prime})L^\prime S;J_k \rangle }\nonumber\\
&=&\left\{ a_0+c_0 +(2S(S+1)-3)(b_0+d_0) \right\}
 \langle (l_\mu l_{\mu^\prime})LS;J_k|\delta({\bm r}-{\bm r}^\prime)|(l_\nu
l_{\nu^\prime})L^\prime S;J_k
 \rangle\nonumber\\
&&+\left\{ a_1+c_1 +(2S(S+1)-3)(b_1+d_1)\right\}
 \langle (l_\mu l_{\mu^\prime})LS;J_k| \kdag{}^2\delta({\bm r}-{\bm r}^\prime)
 +\delta({\bm r}-{\bm r}^\prime){\bm k}^2 |(l_\nu l_{\nu^\prime})L^\prime S;J_k
\rangle \nonumber\\
&&+\left\{ a_2+c_2+(2S(S+1)-3)(b_2+d_2) \right\}
 \langle (l_\mu l_{\mu^\prime})LS;J_k| \kdag \cdot \delta({\bm r}-{\bm r}^\prime)
 {\bm k} |(l_\nu l_{\nu^\prime})L^\prime S;J_k \rangle \nonumber\\
&&+\left\{ a_3+c_3+(2S(S+1)-3)(b_3+d_3) \right\}
\langle (l_\mu l_{\mu^\prime})LS;J_k|\rho_{00}^\alpha({\bm r})\delta({\bm r}-{\bm
r}^\prime)
         |(l_\nu l_{\nu^\prime})L^\prime S;J_k \rangle \nonumber\\
&&+2e_3 \langle (l_\mu l_{\mu^\prime})LS;J_k|\rho_{10}({\bm
r})\rho_{00}^{\alpha-1}({\bm r})
 \delta({\bm r}-{\bm r}^\prime) |(l_\nu l_{\nu^\prime})L^\prime S;J_k \rangle
 \times
\left\{ \begin{array}{cc} (-1) , & {\rm proton} \\ 1 , & {\rm
neutron}\end{array}\right.  \nonumber\\
&&+f_3 \langle (l_\mu l_{\mu^\prime})LS;J_k|\rho^2_{10}({\bm
r})\rho_{00}^{\alpha-2}({\bm r})
 \delta({\bm r}-{\bm r}^\prime) |(l_\nu l_{\nu^\prime})L^\prime S;J_k \rangle
 \nonumber\\
&&+(a_4+c_4)
 \langle (l_\mu l_{\mu^\prime})LS;J_k|({\bm \sigma}+{\bm \sigma}^\prime)\cdot
 \kdag \times \delta({\bm r}-{\bm r}^\prime){\bm k}
 |(l_\nu l_{\nu^\prime})L^\prime S;J_k \rangle ,
\label{v_skyrme_proton_proton}
\end{eqnarray}
\noindent
ii) proton-neutron  matrix elements:
\begin{eqnarray}
\lefteqn{ \langle (l_\mu l_{\mu^\prime})LS;J_k|
\bar{V}^{\text{eff}}_{\text{Skyrme}}
|(l_\nu l_{\nu^\prime})L^\prime S;J_k
\rangle }\nonumber\\
&=& \{ a_0-c_0+(2S(S+1)-3)(b_0-d_0) \} \langle (l_\mu
l_{\mu^\prime})LS;J_k|\delta({\bm r}-{\bm r}^\prime)
|(l_\nu l_{\nu^\prime})L^\prime S;J_k \rangle\nonumber\\
&&+\left\{ a_1-c_1 +(2S(S+1)-3)(b_1-d_1) \right\}\langle (l_\mu
l_{\mu^\prime})LS;J_k|
\kdag{}^2\delta({\bm r}-{\bm r}^\prime)
 +\delta({\bm r}-{\bm r}^\prime){\bm k}^2 |(l_\nu l_{\nu^\prime})L^\prime S;J_k
\rangle \nonumber\\
&&+\left\{ a_2-c_2 + (2S(S+1)-3)(b_2-d_2) \right\}
\langle (l_\mu l_{\mu^\prime})LS;J_k| \kdag \cdot\delta({\bm r}-{\bm r}^\prime)
 {\bm k} |(l_\nu l_{\nu^\prime})L^\prime S;J_k \rangle \nonumber\\
&&+\left\{ a_3-c_3 +(2S(S+1)-3)(b_3-d_3) \right\}
 \langle (l_\mu l_{\mu^\prime})LS;J_k|\rho_{00}^\alpha({\bm r})\delta({\bm r}-{\bm
r}^\prime)
         |(l_\nu l_{\nu^\prime})L^\prime S;J_k \rangle \nonumber\\
&&+f_3 \langle (l_\mu l_{\mu^\prime})LS;J_k|\rho^2_{10}({\bm
r})\rho_{00}^{\alpha-2}({\bm r})
 \delta({\bm r}-{\bm r}^\prime) |(l_\nu l_{\nu^\prime})L^\prime S;J_k \rangle
 \nonumber\\
&&+(a_4-c_4) \langle (l_\mu l_{\mu^\prime})LS;J_k|({\bm \sigma}+{\bm
\sigma}^\prime)\cdot
 \kdag \times \delta({\bm r}-{\bm r}^\prime){\bm k}
 |(l_\nu l_{\nu^\prime})L^\prime S;J_k \rangle .
\label{v_skyrme_proton_neutron}
\end{eqnarray}
We use the canonical (and real) radial wave functions $R_{\mu}(r)$, the
angular wave functions $Y_{l_\mu l_\mu^z}(\Omega)$, and the
spin wave functions
to write the nontrivial matrix elements included in Eqs.\
(\ref{v_skyrme_proton_proton}) and (\ref{v_skyrme_proton_neutron}) as
\begin{eqnarray}
\lefteqn{ \langle (l_\mu l_{\mu^\prime})LS;J_k|\delta({\bm r}-{\bm r}^\prime)|
(l_\nu l_{\nu^\prime})L^\prime S;J_k \rangle }\nonumber\\
&=& \int dr\:r^2 R_\mu(r) R_{\mu^\prime}(r) R_\nu(r) R_{\nu^\prime}(r)
 \delta_{LL^\prime}
 \int d\Omega \left[ Y_{l_\mu}(\Omega)
Y_{l_{\mu^\prime}}(\Omega)\right]^\ast_{L0}
\left[ Y_{l_\nu}(\Omega) Y_{l_{\nu^\prime}}(\Omega)\right]_{L0} , \nonumber\\
\label{matrix_element_delta}
\end{eqnarray}
\begin{equation}
 \int d\Omega \left[ Y_{l_\mu}(\Omega)
Y_{l_{\mu^\prime}}(\Omega)\right]^\ast_{L0}
 \left[ Y_{l_\nu}(\Omega) Y_{l_{\nu^\prime}}(\Omega)\right]_{L0}
= \frac{1}{4\pi}\hat{l_\mu} \hat{l_{\mu^\prime}} \hat{l_\nu}
\hat{l_{\nu^\prime}}
\left(\begin{array}{ccc}l_\mu & l_{\mu^\prime} & L \\ 0 & 0 &
0\end{array}\right)
\left(\begin{array}{ccc}l_\nu & l_{\nu^\prime} & L \\ 0 & 0 &
0\end{array}\right) ,
\end{equation}
\begin{eqnarray}
\lefteqn{
 \langle (l_\mu l_{\mu^\prime})LS;J_k| \delta({\bm r}-{\bm r}^\prime){\bm k}^2
|(l_\nu l_{\nu^\prime})L^\prime S;J_k \rangle
}\nonumber\\
&=& -\frac{1}{4}\delta_{LL^\prime}
 \int d\Omega \left[ Y_{l_\mu}(\Omega)
Y_{l_{\mu^\prime}}(\Omega)\right]^\ast_{L0}
 \left[ Y_{l_\nu}(\Omega) Y_{l_{\nu^\prime}}(\Omega)\right]_{L0}
 \nonumber\\
&&\times \int dr\:r^2 R_\mu(r) R_{\mu^\prime}(r)
\left\{ \left[ \left(
\frac{d^2}{dr^2}+\frac{2}{r}\frac{d}{dr}-\frac{l_\nu(l_\nu+1)}{r^2}
\right)R_\nu(r)\right]R_{\nu^\prime}(r)\right.\nonumber\\
&&\left.+R_\nu(r)\left[ \left( \frac{d^2}{dr^2}+\frac{2}{r}\frac{d}{dr}
 -\frac{l_{\nu^\prime}(l_{\nu^\prime}+1)}{r^2}\right)R_{\nu^\prime}(r)
 \right]\right\}\nonumber\\
&&-\sum_{\Delta l_\nu=0,1} \sum_{\Delta l_{\nu^\prime}=0,1}
 \frac{1}{2}\sqrt{(l_\nu+\Delta l_\nu)(l_{\nu^\prime}+\Delta l_{\nu^\prime})}
 \left\{\begin{array}{ccc}l_{\nu^\prime}-1+2\Delta l_{\nu^\prime} &
 l_{\nu}-1+2\Delta l_{\nu} & L \\
 l_\nu  & l_{\nu^\prime} & 1 \end{array}\right\} \nonumber\\
&&\times\int dr\:r^2 R_\mu(r) R_{\mu^\prime}(r)
\left\{ (l_\nu+1-\Delta l_\nu)\frac{R_\nu(r)}{r}
 +(-)^{\Delta l_\nu}\frac{dR_\nu(r)}{dr}\right\}\nonumber\\
&&\times \left\{ (l_{\nu^\prime}+1-\Delta
l_{\nu^\prime})\frac{R_{\nu^\prime}(r)}{r}
 +(-)^{\Delta l_{\nu^\prime}}\frac{dR_{\nu^\prime}(r)}{dr}\right\}\nonumber\\
&&\times
 \delta_{LL^\prime}
 \int d\Omega \left[ Y_{l_\mu}(\Omega)
Y_{l_{\mu^\prime}}(\Omega)\right]^\ast_{L0}
 \left[ Y_{ l_{\nu^\prime}-1+2\Delta l_{\nu^\prime} }(\Omega)
 Y_{l_\nu-1+2\Delta l_\nu} (\Omega)\right]_{L0} ,
\end{eqnarray}
\begin{equation}
\langle (l_\mu l_{\mu^\prime})LS;J_k| \kdag \cdot\delta({\bm r}-{\bm r}^\prime)
 {\bm k} |(l_\nu l_{\nu^\prime})L^\prime S;J_k \rangle
= -\frac{\sqrt{3}}{\hat{L}}\delta_{LL^\prime}
 \langle (l_\mu
l_{\mu^\prime})L||(\kdag \cdot
\delta({\bm r}-{\bm r}^\prime) {\bm k})_0||
(l_\nu l_{\nu^\prime})L\rangle ,
\label{t2}
\end{equation}
\begin{eqnarray}
\lefteqn{
\langle (l_\mu l_{\mu^\prime})LS;J_k|i({\bm \sigma}+{\bm \sigma}^\prime)\cdot
 \kdag \times \delta({\bm r}-{\bm r}^\prime){\bm k}
 |(l_\nu l_{\nu^\prime})L^\prime S;J_k \rangle }\nonumber\\
&=& (-)^{1+L^\prime+J_k}4 \sqrt{3}
 \left\{ \begin{array}{ccc} 1 & L^\prime & J_k \\ L & 1 & 1 \end{array}\right\}
 \delta_{S1}
 \langle (l_\mu
l_{\mu^\prime})L|| (\kdag \cdot \delta({\bm r}-{\bm r}^\prime){\bm k})_1||
(l_\nu l_{\nu^\prime})L'\rangle .
\label{spin_orbit}
\end{eqnarray}
The square brackets around products of spherical harmonics and the
parentheses surrounding products of operators indicate
angular-momentum coupling.

To evaluate Eqs.\ (\ref{t2}) and (\ref{spin_orbit}), one can use
\begin{eqnarray}
\lefteqn{ \langle (l_\mu
l_{\mu^\prime})L||( \kdag \cdot
\delta({\bm r}-{\bm r}^\prime){\bm k})_I||
(l_\nu l_{\nu^\prime})L'\rangle }\nonumber\\
&=&\left[\frac{1}{4}\sum_{\Delta l_\mu=0,1} \sum_{\Delta l_\nu=0,1}
\sum_{l_{\mu\mu^\prime}}
 \int dr\:r^2 \left\{ (l_\mu+1-\Delta l_\mu)\frac{R_\mu(r)}{r}
 +(-)^{\Delta l_\mu}\frac{dR_\mu(r)}{dr} \right\} R_{\mu^\prime}(r)
\right.\nonumber\\
&& \times \left\{ (l_\nu+1-\Delta l_\nu)\frac{R_\nu(r)}{r}
 +(-)^{\Delta l_\nu}\frac{dR_\nu(r)}{dr} \right\} R_{\nu^\prime}(r) \nonumber\\
&&\times\sqrt{l_\mu+\Delta l_\mu}
        \sqrt{l_\nu+\Delta l_\nu}
        \sqrt{2l_\mu+4\Delta l_\mu-1}
              \hat{ l_{\mu^\prime} }
        \sqrt{2l_\nu+4\Delta l_\nu-1}
              \hat{ l_{\nu^\prime} }
 \nonumber\\
&&\times \frac{1}{4\pi}\hat{l}^2_{\mu\mu^\prime}
              \hat{L}
              \hat{L}^\prime
              \hat{I}
 (-)^{l_\mu+l_{\mu^\prime}+L+I+1}
 \left\{ \begin{array}{ccc}I & L^\prime & L \\ l_{\mu\mu^\prime} & 1 & 1
\end{array} \right\}
 \left\{ \begin{array}{ccc}l_{\mu\mu^\prime} & L & 1 \\ l_\mu & l_\mu+2\Delta
l_\mu-1 &
 l_{\mu^\prime} \end{array} \right\}
\nonumber\\
&&\times
 \left\{ \begin{array}{ccc}l_{\mu\mu^\prime} & L^\prime & 1 \\ l_\nu &
l_\nu+2\Delta l_\nu-1 &
 l_{\nu^\prime} \end{array} \right\}
 \left( \begin{array}{ccc}l_\mu +2\Delta l_\mu-1 & l_{\mu^\prime} &
l_{\mu\mu^\prime} \\
 0 & 0 & 0 \end{array} \right)
\nonumber\\
&&\times
 \left.
 \left( \begin{array}{ccc}l_\nu +2\Delta l_\nu-1 & l_{\nu^\prime} &
l_{\mu\mu^\prime} \\
 0 & 0 & 0 \end{array} \right)
 \right]
 -(-)^{l_\nu+l_{\nu^\prime}+L^\prime}[\nu \leftrightarrow \nu^\prime]
 -(-)^{l_\mu+l_{\mu^\prime}+L}[\mu \leftrightarrow \mu^\prime]
\nonumber\\
&&
 +[\mu \leftrightarrow \mu^\prime {\rm \ and\ } \nu \leftrightarrow \nu^\prime ]
,
\end{eqnarray}
where for reduced matrix elements we have used the convention
\begin{equation}
\langle LL_z |\hat{O}_{lm}|L^\prime L^\prime_z \rangle =
\frac{1}{\hat{L}}
\langle L^\prime L^\prime_z lm|L L_z \rangle \langle L||\hat{O}_l||L^\prime
\rangle ,
\end{equation}
and made the abbreviation
\begin{eqnarray}
&&[A_{\mu\mu^\prime\nu\nu^\prime}]  -(-)^{l_\nu+l_{\nu^\prime}+L^\prime}[\nu \leftrightarrow \nu^\prime]
 -(-)^{l_\mu+l_{\mu^\prime}+L}[\mu \leftrightarrow \mu^\prime]
 + [\mu \leftrightarrow \mu^\prime {\rm \ and\ }
    \nu \leftrightarrow \nu^\prime] \nonumber\\
&\equiv&
A_{\mu\mu^\prime\nu\nu^\prime} -
(-)^{l_\nu+l_{\nu^\prime}+L^\prime} A_{\mu\mu^\prime\nu^\prime\nu} -
(-)^{l_\mu+l_{\mu^\prime}+L}A_{\mu^\prime\mu\nu\nu^\prime} +
A_{\mu^\prime\mu\nu^\prime\nu} .
\end{eqnarray}

Eq.\ (\ref{matrix_element_delta}), modified to include additional factors in
the radial integral, can also be used (together with the subsequent equations)
to evaluate the matrix elements of the terms involving $\rho_{00}^\alpha({\bm
r})$ in $ \bar{V}^{\rm eff}_{\rm Skyrme}$, the
Coulomb-exchange interaction, and the contributions of the pairing functional
to the effective ph, pp, and 3p1h interactions.
The Coulomb-direct term can be evaluated in a similar but slightly more
complicated way, via a multipole expansion.

In the main part of this paper we used the Skyrme
functional SkM$^\ast$, which is usually parameterized as in interaction in
terms  of coefficients $t_0,t_1,t_2,t_3,x_0,x_1,x_2, x_3,$ and $W_0$.
The relations between these coefficients and those used here, if no terms
are neglected, are \cite{dob95,per04}
\begin{equation}
\begin{array}{ll}
%
C_0^{\rho} =  \frac{3}{8} t_0
             +\frac{3}{48}t_3 \, \rho_{00}^\alpha, &
C_1^{\rho} = -\frac{1}{4} t_0(\frac{1}{2}+x_0)
             -\frac{1}{24}t_3(\frac{1}{2}+x_3)\, \rho_{00}^\alpha, \\
C_0^{s}    = -\frac{1}{4} t_0(\frac{1}{2}-x_0)
             -\frac{1}{24}t_3(\frac{1}{2}-x_3)\, \rho_{00}^\alpha,  &
C_1^{s}    = -\frac{1}{8} t_0
             -\frac{1}{48}t_3\, \rho_{00}^\alpha,  \\
C_0^{\tau} =  \frac{3}{16}t_1 +\frac{1}{4}t_2(\frac{5}{4}+x_2), &
C_1^{\tau} = -\frac{1}{8} t_1(\frac{1}{2}+x_1) +\frac{1}{8}t_2(\frac{1}{2}+x_2), \\
C_0^{T}    = -\frac{1}{8} t_1(\frac{1}{2}-x_1)+\frac{1}{8}t_2(\frac{1}{2}+x_2), &
C_1^{T}    = -\frac{1}{16}t_1+\frac{1}{16}t_2,  \\
C_0^{\Delta\rho} = -\frac{9}{64}t_1+\frac{1}{16}t_2(\frac{5}{4}+x_2),  &
C_1^{\Delta\rho} =  \frac{3}{32}t_1(\frac{1}{2}+x_1)+\frac{1}{32}t_2(\frac{1}{2}+x_2),  \\
C_0^{\Delta s}   =  \frac{3}{32}t_1(\frac{1}{2}-x_1)+\frac{1}{32}t_2(\frac{1}{2}+x_2) , \qquad   &
C_1^{\Delta s}   =  \frac{3}{64}t_1+\frac{1}{64}t_2,  \\
C_0^{\nabla J} = -\frac{3}{4} W_0,  &
C_1^{\nabla J} = -\frac{1}{4} W_0 .
\end{array}
\end{equation}
In the HF fits that originally determined the SkM$^\ast$ parameters, the effects of $C_t^T$ (the ``$J^2$
terms'') were
neglected because of technical difficulties.  These terms have often been
included in subsequent RPA calculations.  To maintain self consistency here, 
we have
set them to zero both
in the HFB calculation and in the QRPA.

\end{document}